\documentclass[lettersize,journal]{IEEEtran}
\usepackage{amsmath,amsfonts}
\usepackage{algorithmic}
\usepackage{algorithm}
\usepackage{array}
\usepackage[caption=false,font=normalsize,labelfont=sf,textfont=sf]{subfig}
\usepackage{textcomp}
\usepackage{stfloats}
\usepackage{url}
\usepackage{verbatim}
\usepackage{graphicx}
\usepackage[export]{adjustbox}
\usepackage{cite}
\usepackage{booktabs}
\usepackage{multirow}
\usepackage{xcolor}
\newcommand{\eg}{e.\,g.,\ }
\newcommand{\ie}{i.\,e.,\ }

\usepackage{threeparttable}
\usepackage{tablefootnote}

\newcommand{\editor}[1]{{\color{black}#1}}     
\newcommand{\reviewone}[1]{{\color{black}#1}}   
\newcommand{\reviewtwo}[1]{{\color{black}#1}}   
\newcommand{\reviewthree}[1]{{\color{black}#1}}

\begin{document}

% \title{PABench: A Unified Evaluation Benchmark for Paralinguistic Analysis}
% ParaLBench: An Extensive Benchmark for Computational Paralinguistics Using Acoustic Foundation Models
\title{ParaLBench: A Large-Scale Benchmark \\ for Computational Paralinguistics \\ over Acoustic Foundation Models}

% \name{Zixing Zhang$^1$, Weixiang Xu$^1$, Zhongren Dong$^1$, Tao Pang$^1$, Liyizhe Peng$^1$, Runming Wang$^2$,\\ \textit{Huan Zhao$^1$, Bj\"orn W.\ Schuller$^3$}

\author{Zixing Zhang,~\IEEEmembership{Senior Member,~IEEE}, Weixiang Xu, Zhongren Dong,~\IEEEmembership{Student Member,~IEEE},\\Kanglin Wang, Yimeng Wu, Jing Peng, Runming Wang, Dong-Yan Huang,~\IEEEmembership{Senior Member,~IEEE}
        % <-this % stops a space

\thanks{The work was supported by the Guangdong Basic and Applied Basic Research Foundation under Grant Number 2024A1515010112. (Corresponding authors: Zixing~Zhang)}
\thanks{Z.~Zhang, W.~Xu, Z.~Dong, K.~Wang, and J.~Peng are with the College of Computer Science and Electronic Engineering, Hunan University, Changsha 410082, China; Z.~Zhang is also with the Shenzhen Research Institute, Hunan University, Shenzhen 518000, China. e-mail: \{zixingzhang, xuweixiang, zrdong, wangkanglin, yimengwu, johanna\}@hnu.edu.cn}% <-this % stops a space
\thanks{R.~Wang is with the School of Information Science and Engineering, Hunan Normal University, Changsha 410081, China. e-mail: runminwang@hunnu.edu.cn}
\thanks{D.~Huang is with the UBTECH Robotics Corp, Shenzhen 518071, China. e-mail: dongyan.huang@ubtrobot.com}
}

% \markboth{IEEE Transactions on Affective Computing}%
{}


% \markboth{IEEE Transactions on Affective Computing}%


\maketitle

\begin{abstract}

Computational paralinguistics (ComParal) aims to develop algorithms and models to automatically detect, analyze, and interpret non-verbal information from speech communication, \eg emotion, health state, age, and gender. Despite its rapid progress, it heavily depends on \reviewtwo{sophisticatedly} designed models given specific paralinguistic tasks. Thus, the heterogeneity and diversity of ComParal models largely prevent the realistic implementation of ComParal models. Recently, with the advent of acoustic foundation models because of self-supervised learning, developing more generic models that can efficiently perceive a plethora of paralinguistic information has become an active topic in speech processing. However, it lacks a unified evaluation framework for a fair and consistent performance comparison. To bridge this gap, we conduct a large-scale benchmark, namely \textit{ParaLBench}, which concentrates on standardizing the evaluation process of diverse paralinguistic tasks, \editor{including critical aspects of affective computing such as emotion recognition and emotion dimensions prediction,} over different acoustic foundation models. This benchmark contains \textcolor{black}{ten} datasets with \textcolor{black}{thirteen} distinct paralinguistic tasks, covering short-, medium- and long-term characteristics. Each task is carried out on 14 acoustic foundation models under a unified evaluation framework, which allows for an unbiased methodological comparison and offers a grounded reference for the ComParal community. 
Based on the insights gained from ParaLBench, we also point out potential research directions, \ie the cross-corpus generalizability, to propel ComParal research in \textcolor{black}{the} future. The code associated with this study will be available to foster the transparency and replicability of this work for succeeding researchers.

\end{abstract}


\section{Introduction}

\begin{figure}[t]
  \centering
  \includegraphics[width=1 \columnwidth,trim=120 230 60 150, clip]{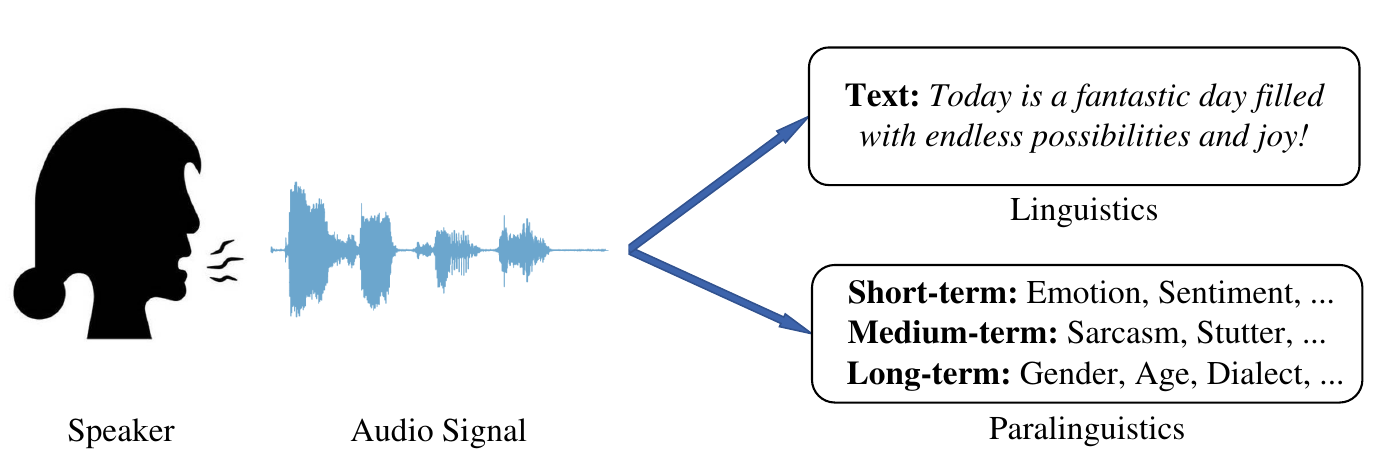}
  % \caption{Paralinguistics, also known as ``alongside linguistics", aims to study all other information from speech in addition to linguistics, such as emotion, age, gender, health status, etc.}
    \caption{Compared to linguistics that studies the text content, \textit{paralinguistics} analyzes all other information from speech, such as emotion, age, gender, and health status.}
  \label{fig:paralinguistic}
\end{figure}

\IEEEPARstart{C}{omputational} paralinguistics (ComParal) refers to the interdisciplinary field of study that focuses on the computational analysis and processing of non-verbal elements of communication in human speech. \editor{For example, affective computing, a core aspect of ComParal, focuses specifically on the recognition and understanding of emotions from speech, making it a major task in this domain.} These non-verbal elements, also known as paralinguistic features, include aspects, such as tone of voice, intonation, pitch, loudness, speech rate, pauses, and emotional expression. They are crucial for conveying the authentic intent of speakers beyond syntax and vocabulary~\cite{schuller2013paralinguistics}, and thus play an important role in human-computer interaction~\cite{9257201}, healthcare~\cite{10446795}, public security~\cite{verhagen2020covid}, and others. The difference between linguistic and paralinguistic information is illustrated in \textcolor{black}{Figure}~\ref{fig:paralinguistic}. The goal of ComParal is to develop algorithms and systems that can automatically detect, interpret, and analyze these paralinguistic cues for various applications, including emotion recognition, speaker identification, and speech therapy.

Over the past decades, ComParal has made considerable progress~\cite{8682896,koenecke2020racial,8462579}, partially thanks to the series of ComParE challenges~\cite{Schuller09-IS, Schuller23-ACM}. 
However, the disparate selection of features across these algorithms has resulted in inconsistent performance across various tasks, thereby posing a significant challenge in assessing the effectiveness of each algorithm.
Early studies~\cite{powell2022deep,wang2020speech,kshirsagar2022quality} primarily relied on small-scale handcrafted features, such as conventional Mel-Frequency Cepstral Coefficients (MFCC) and log Mel-Filter Banks (log-Mel). Later on, large-scale sophisticatedly designed handcrafted features extracted by openSMILE~\cite{eyben2010opensmile} and openXBOW~\cite{schmitt2017openxbow} toolkits were widely utilized in this field. The openSMILE applies many statistical functionals (\eg mean and variance), \textcolor{black}{while} the openXBOW explores a bag-of-word strategy, on top of a set of \editor{low-level descriptors (LLD, \eg MFCC and pitch)}, resulting in more than thousands of acoustic features over one audio segment.
% to a set of low-level descriptors (LLD, \eg MFCC and pitch), resulting in more than thousands of acoustic features over one audio segment. In contrast, the openXBOW~\cite{} explores the bag-of-word strategy in the speech domain by taking vector quantization to discrete continuous acoustic features. 

Recently, with the development of self-supervised learning (SSL), more and more acoustic foundation models have been proposed, aiming to extract more general and efficient acoustic representations. These models include, but are not limited to, wav2vec~\cite{schneider2019wav2vec}, HuBERT~\cite{hsu2021hubert}, WavLM~\cite{chen2022wavlm}, data2vec~\cite{hsu2021hubert}, and Whisper~\cite{radford2022whisper}. Although these models were initially proposed and evaluated for automatic speech recognition (ASR), increasing studies have demonstrated their capability in diverse ComParal applications with or without fine-tuning or adaptation paradigms~\cite{chen2023speechformer++,10446795,yang2024large}. Therefore, it is necessary to evaluate these typical and widely used features across varied ComParal tasks and to provide a full picture of the performance of these acoustic feature extractors in ComParal. This is crucial to explore their potential capabilities that go beyond the transcription of linguistic content from speech. 

In addition, the inconsistencies in evaluation metrics and data partitioning further increase the difficulty of a fair performance comparison. For example, the MSP-Podcast dataset~\cite{msp-podcast}, which is extensively used in emotional domain studies, encounters issues due to continuous updates and its collection from diverse podcast sources. The use of different versions by various researchers complicates the performance comparison. Thus, it is essential to standardize the evaluation framework for an unbiased performance comparison. 

Moreover, with the advent of large multimodal models, such as GPT-4o and Gemini~\cite{team2023gemini}, it is increasingly interesting how these models perform in ComParal to evaluate their \textcolor{black}{universality} 
% university 
and efficiency of paralinguistic perception, and even the capability of cognition. Before these studies, it lacks a fundamental benchmark to assess their potential comprehensively.

To address the inconsistency challenges and bridge the thorough performance comparison gap, we introduce a benchmark for ComParal, namely \textit{ParaLBench}, with the datasets, the tasks, and the foundational acoustic models that are widely investigated in this field. Besides, 
we standardize the evaluation metrics and employ a consistent data partitioning strategy. 
Therefore, ParaLBench offers a comprehensive understanding of ComParal, and tends to establish fair and productive comparisons among diverse computational approaches and to foster systematic advancement in paralinguistic analysis. 
%Besides
\textcolor{black}{Moreover}, we particularly delve into the pros and cons of foundational acoustic models for various paralinguistic datasets and tasks.

Overall, the contributions of this article are summarized as follows:
\begin{itemize}
    \item We establish a unified and state-of-the-art benchmark over large-scale ComParal tasks, including feature extraction, data partition, and evaluation metric selection. To the best of our knowledge, this is the most comprehensive and standard work in the ComParal field to date. This benchmark provides an unbiased assessment ground and thus can facilitate the model development of ComParal.
    \item We thoroughly evaluate thirteen diverse paralinguistic tasks over 14 popularly used acoustic foundation models. By empirically investigating the performance of these models in ComParal, it greatly helps us understand the generality, efficiency, and robustness of current acoustic foundation models, and find out their drawbacks that can be further optimized in future work. 
\end{itemize}

% This paper is organized as follows: 
% Section \ref{sec:Related Works} outlines advancements in ComParal within the field of machine learning, introduces commonly used acoustic features in analysis tasks, and concludes with a discussion on related benchmarks in speech and paralinguistics.
% Section \ref{sec:The Benchmark Methods} describes the corpora used in this study, specific assessment tasks, acoustic foundation models, and standardized metrics.
% Section \ref{sec: The ComParal Corpus and Taxonomy} details the benchmark models and their parameter settings. 
% Section \ref{sec:Benchmark} establishes ParaLBench, presents extensive empirical work, and comprehensively analyzes the results.
% We conclude the paper by summarizing our key findings and exploring potential future research directions.

The remainder of the present paper is organized as follows: 
\textcolor{black}{First}, we briefly introduce the advancements of ComParal in Section \ref{sec:Related Works}. 
We then describe a unified evaluation framework for our ParaLBench experiments, including the acoustic foundation models, the selected paralinguistic tasks, and the evaluation metrics in Section~\ref{sec:The Benchmark Methods}. 
In Section~\ref{sec:experiment}, we present the employed datasets as well as the unified experimental setups for all experiments. Corresponding experimental results and detailed analysis are followed in Section~\ref{sec:result}. 
% We finally
\textcolor{black}{Finally,} we conclude the work by summarizing key findings and pointing out potential research directions in Section~\ref{sec:Conclusion}.

% In Section \ref{sec:Related Works}, we will first provide a brief overview of computational paralinguistics and recent advancements in the field, followed by an introduction to some commonly used features in the speech and language processing domains. Section \ref{sec:The Benchmark Methods} will cover the benchmark models, parameter settings, acoustic base models (features), and standardized metrics used in this benchmark. In Section \ref{sec: The Computational Paralinguistics Corpus and Taxonomy}, we will delve into a detailed description of the datasets used in this study and categorize different paralinguistic tasks according to ~\cite{schuller2013computational}. Subsequently, in Section \ref{sec:Benchmark}, we establish ParaLBench and conduct extensive experiments and result analyses. Finally, we conclude the paper by summarizing our findings and discussing some promising research directions.

\section{Related Works}
\label{sec:Related Works}
% This paper primarily focuses on the domain of paralinguistic analysis. In this section, we initially provide an overview of the tasks and applications involved in paralinguistic analysis, followed by a summary of recent major research efforts, and finally, a brief exposition of some commonly used features in the field of speech processing.

% This section focuses on the domain of paralinguistic analysis, outlining its key aspects. It presents an overview of relevant tasks and applications, summarizes recent major research advancements, and discusses commonly utilized features in speech processing. Additionally, the section highlights the role of benchmarking in improving evaluation methodologies within this field.

This section presents an overview of relevant tasks and applications in ComParal, summarizes its major advancements, and discusses commonly utilized features. Additionally, it highlights the role of benchmarking in improving evaluation methodologies in this field.

% \subsection{Computational Paralinguistics}
\subsection{Foundations of Computational Paralinguistics}
% 简单描述一下什么是副语言，随后描述一下哪些任务属于副语言分析
% Paralinguistic, also known as "alongside linguistics"~\cite{schuller2013computational, schuller2013paralinguistics, velichko2022complex}, is the process of studying non-verbal elements beyond speech aimed at understanding how these elements impact communication and information transmission. These elements encompass intonation~\cite{liberman1975intonational, tang2023qi}, speaking speed~\cite{sommers2020age, knowlton2006influence}, volume~\cite{page1978effect, knowlton2006influence}, as well as speaker characteristics such as gender~\cite{meena2013gender, ali2012gender, alamsyah2020speech}, age~\cite{kaya2017emotion, sedaghi2009comparative, radha2024automatic}, emotions~\cite{chen2023speechformer++, chen2022key, 10447044}, and health status~\cite{ye2021health, hassan2020covid}.
Paralinguistic analysis, a component of the broader discipline known as ``alongside linguistics''~\cite{schuller2013computational, schuller2013paralinguistics, velichko2022complex}, investigates non-verbal elements that complement spoken language, aiming to elucidate their role in enhancing communication and information transmission.
These elements encompass intonation~\cite{liberman1975intonational, tang2023qi}, speech rate~\cite{sommers2020age, knowlton2006influence}, and volume~\cite{page1978effect, knowlton2006influence}, which convey subtleties and encapsulate speaker attributes such as gender~\cite{meena2013gender, ali2012gender, alamsyah2020speech}, age~\cite{kaya2017emotion,sedaghi2009comparative, radha2024automatic}, emotion~\cite{chen2023speechformer++, chen2022key, 10447044}, and health status~\cite{ye2021health, hassan2020covid}.
\reviewtwo{Paralinguistic analysis} thus enriches our understanding of human discourse, with diverse applications such as emotion recognition~\cite{chen2023speechformer++, chen2022key, 10447044}, detecting deceptive communication~\cite{constancio2023deception}, speaker identification~\cite{jahangir2021speaker, bai2021speaker}, enhancing human-computer interaction~\cite{alnuaim2022human, ramakrishnan2013speech}, and aiding healthcare diagnostics~\cite{ye2021health, hassan2020covid}.

However, synthesizing this multifaceted data into coherent frameworks for comprehensive interpretation remains a challenge. The development of ParaLBench is motivated by the need to address these complexities and the desire to create a unified platform that leverages the diverse aspects of paralinguistics for both scholarly investigation and practical application.

% \subsection{Paralinguistic on Machine Learning}
\subsection{Machine Learning in Paralinguistic Analysis}
% 根据副语言任务进行划分
% 从传统机器学习方法开始介绍，随后介绍深度学习方法
% The speech signal contains information about the speaker's identity, emotional state, language habits, and so forth. These features can be analyzed and recognized through aspects such as intonation, speech rate, and speech quality. Therefore, researchers are dedicated to exploring more efficient methods for paralinguistic analysis~\cite{chen2023speechformer++, 10446795}.

Speech signals are rich in information pertaining to the speaker's identity, emotional state, and language habits, which can be discerned through analysis of intonation, speech rate, and timbre. The evolution of machine learning has propelled paralinguistic analysis from rudimentary rule-based and statistical methods to sophisticated deep learning techniques~\cite{chen2023speechformer++, 10446795}.

% During the early stages of machine learning, paralinguistic analysis mainly relied on rule-based symbolic approaches~\cite{wang2002combination, de1987rule} and statistical learning~\cite{tokuda1995speech, schuller2003hidden, yang2016decision, reichl2000robust, poon2014acoustic}. Researchers attempted to use expert rules or statistical models to analyze non-linguistic elements in speech, such as emotion and healthy state. For example, in ~\cite{schuller2003hidden}, low-level instantaneous features were utilized to classify speech emotions using continuous Hidden Markov Models. Similarly, decision trees were employed for depression classification in ~\cite{yang2016decision}. However, these methods were limited by the complexity of manually crafting rules and the constraints of statistical models, making it difficult to handle complex speech signals.

Initially, paralinguistic analysis depended on symbolic techniques and statistical models, guided by expert-designed rules, to interpret diverse non-linguistic features~\cite{wang2002combination, de1987rule,tokuda1995speech, schuller2003hidden, yang2016decision, reichl2000robust, poon2014acoustic}. Nonetheless, these methods were constrained in their ability to manage the intricate nature of speech data. Subsequent computational advancements, particularly through deep learning techniques such as convolutional neural networks (CNNs)~\cite{guo2021representation, seneviratne2020generalized, makiuchi2021speech} and recurrent neural networks (RNNs)~\cite{mirsamadi2017automatic, dumpala2021significance, nitisara2020speech}, have exhibited remarkable proficiency in discerning significant patterns from speech signals.

The introduction of Transformers~\cite{vaswani2017attention, devlin2018bert, radford2018improving}, featuring their self-attention mechanisms, marked a significant milestone, as they enhanced speech analysis by effectively handling long-range temporal dependencies. Innovations like SpeechFormer++~\cite{chen2023speechformer++} and HAFFormer~\cite{10446795} exemplify this progress, tackling computational challenges and refining the analysis of paralinguistic tasks.

Notwithstanding these advancements, the field continues to grapple with challenges related to generalization, data availability, and model complexity. ParaLBench has emerged as an essential tool to confront these issues, providing a systematic platform for assessing model capabilities and performance in paralinguistic tasks.

% \subsection{Acoustic Features}
\subsection{Acoustic Features for Paralinguistic Analysis}
% 从手工特征开始，随后介绍SSL特征

% The paralinguistic information is crucial for understanding various paralinguistic analysis tasks, such as emotion, gender, and age. Early research efforts adapted to these tasks by designing handcraft features. For instance, MFCC~\cite{wu2005improved, zhao2013analyzing} (Mel Frequency Cepstral Coefficients) is one of the most common acoustic features, extracted by applying Mel filters and discrete cosine transform to speech signals. eGeMAPS~\cite{neumann2017attentive} extends the traditional GeMAPS feature set, encompassing additional acoustic features and functionalities, such as fundamental frequency variation and energy in different frequency bands. ComParE-2016~\cite{asci2020machine, edwards2020multiscale}, on the other hand, is a feature set specifically designed for speech emotion recognition tasks, encompassing multiple acoustic and statistical features, along with novel feature engineering methods aimed at enhancing emotion recognition performance. These features are extracted in different ways, hence potentially encoding different paralinguistic information. 

Acoustic features are crucial in paralinguistic analysis, disclosing information about emotion, gender, and age. Historically, research predominantly relied on handcrafted features such as MFCC~\cite{wu2005improved, zhao2013analyzing}. Innovations have broadened the feature repertoire, incorporating more advanced acoustic profiles like \editor{extended Geneva Acoustic Parameter Set (eGeMAPS)}~\cite{neumann2017attentive} and task-specific sets such as ComParE-2016~\cite{asci2020machine, edwards2020multiscale}.

% In recent years, with the emergence of self-supervised learning (SSL) like BERT~\cite{devlin2018bert} and VIT~\cite{dosovitskiy2020image}, learning universal representations of audio from large-scale unlabeled speech corpora has become feasible. For instance, wav2vec~\cite{schneider2019wav2vec} employs a fully convolutional network architecture, taking raw audio as input and optimizing it through a noise contrastive binary classification task. wav2vec 2.0~\cite{baevski2020wav2vec}, an enhanced version of wav2vec, adopts the Transformer as its main architecture and employs a training method called "contrastive predictive coding" to learn speech representations by maximizing the similarity between positive and negative samples. HuBERT~\cite{hsu2021hubert} utilizes an offline clustering step to provide aligned target labels for BERT-like prediction loss, significantly enhancing the quality of learned representations. Due to differences in optimization objectives and model architectures, different self-supervised models may capture distinct paralinguistic information. Therefore, we will mainly explore the performance of different acoustic basis models on different Computational Paralinguistics tasks to provide guidance for subsequent research.

Despite the benefits, traditional handcraft features were not without their limitations, particularly in terms of adaptability to the complex and varied nature of speech signals. The advent of SSL models like BERT~\cite{devlin2018bert} and VIT~\cite{dosovitskiy2020image}, however, heralded a new era in acoustic feature extraction. Contemporary models such as wav2vec~\cite{schneider2019wav2vec} and its successor, wav2vec 2.0~\cite{baevski2020wav2vec}, employ advanced techniques to auto-encode audio input, enhancing the precision and efficiency of paralinguistic analysis. This trend extends to even more recent models like HuBERT~\cite{hsu2021hubert}, which leverage offline clustering to improve alignment and predictive accuracy of speech representations.

The progression from handcrafted to self-supervised learning-based features marks not just methodological innovation but also a practical stride toward robust features capable of generalizing across linguistically varied datasets with reduced manual effort. Accordingly, ParaLBench will conduct a thorough evaluation of these models to guide future research.

\subsection{Benchmark in Speech and Paralinguistic Analysis}
Benchmarks play a pivotal role in standardizing evaluation by harmonizing datasets, methods, and models, thereby propelling research and development. Within the realm of speech processing, numerous benchmarks~\cite{yang2021superb,javed2023indicsuperb,shi2023ml,yang2024large,scheidwasser2022serab,phukan2024paralingual,lian2024merbench} have been instituted to gauge the universality and adaptability of speech representation methods as technology evolves. For instance, the Speech processing Universal PERformance Benchmark (SUPERB)~\cite{yang2021superb} offers a structured framework for systematically assessing the performance of speech representations across an array of tasks, encompassing content understanding, speaker identification, semantic analysis, and paralinguistic feature recognition. Although SUPERB has set a high standard, its predominant reliance on English language corpora limits its global applicability, leading to the emergence of benchmarks like IndicSUPERB~\cite{javed2023indicsuperb}, which focuses on Indian languages, and ML-SUPERB~\cite{shi2023ml}, extending to a broader linguistic spectrum that includes up to 143 languages, significantly enhancing cross-lingual evaluation capabilities and facilitating the improvement of model performance and generalizability in varied linguistic contexts. Additionally, a recent study~\cite{yang2024large} has expanded SUPERB by conducting extensive experiments to evaluate the performance of various SSL models on speech processing tasks, affirming the generalizability and efficacy of SSL models across diverse tasks.

Additionally, beyond the realm of general speech processing tasks, specialized benchmarks have emerged, specifically designed for computational paralinguistic tasks. For example, SERAB~\cite{scheidwasser2022serab} scrutinizes utterance-level Speech Emotion Recognition (SER), with a particular emphasis on the application and generalization potential of diverse feature sets. In contrast to SERAB, which concentrates on features derived from deep neural networks, studies on SSL models, notably the paralinguistic model \textcolor{black}{TRILLsson~\cite{shor2022trillsson}}, have demonstrated their superior efficacy in SER tasks, highlighting TRILLsson's exceptional performance across multiple languages.
%[68]. 
Furthermore, MERBench~\cite{phukan2024paralingual} introduces a novel unified benchmark for multimodal emotion recognition, crafted to assist researchers by offering clear guidelines for both methodological advancement and thorough interaction with multimodal emotional data.

These benchmarks have propelled the field forward. Nevertheless, many have been confined to narrow research domains, such as single-language assessments or specific facets of paralinguistics like emotion recognition, resulting in a lack of comprehensive evaluation platforms that can cover the full range of ComParal tasks. Acknowledging these gaps in evaluation and comparison, ParaLBench has been developed not just as another benchmark, but as a holistic analytical toolkit crafted to evaluate a broad array of paralinguistic activities. ParaLBench's scope encompasses a variety of tasks, including speaker style recognition and accent detection, which are poised to significantly advance research and usher in a new era of progress in ComParal.

\section{ParaLBench framework}
\label{sec:The Benchmark Methods}
% In this section, we first introduce the model structure. Subsequently, existing tasks were classified based on the taxonomy method in~\cite{schuller2013paralinguistics, schuller2013computational}, to facilitate subsequent experimental analysis. Thereafter, we describe the acoustic foundation models assessed in our study. Finally, we elucidate the criteria for evaluation.

In this section, we first describe the unified evaluation model structure, followed by the introduction of two handcrafted acoustic features and 14 widely used acoustic foundation models. Then, we present the paralinguistic tasks assessed in our study, and finally elucidate the criteria for evaluation.

\subsection{Model Structure}
% The objective of ParaLBench is to impartially assess the generalization capability of various acoustic foundational models in performing computational linguistics tasks. Establishing a standardized model architecture is essential for accurately determining an acoustic model's performance without influences from auxiliary model enhancements.

\begin{figure}[!t]
  \centering
  \includegraphics[width=0.8\columnwidth,trim=310 165 260 110, clip]{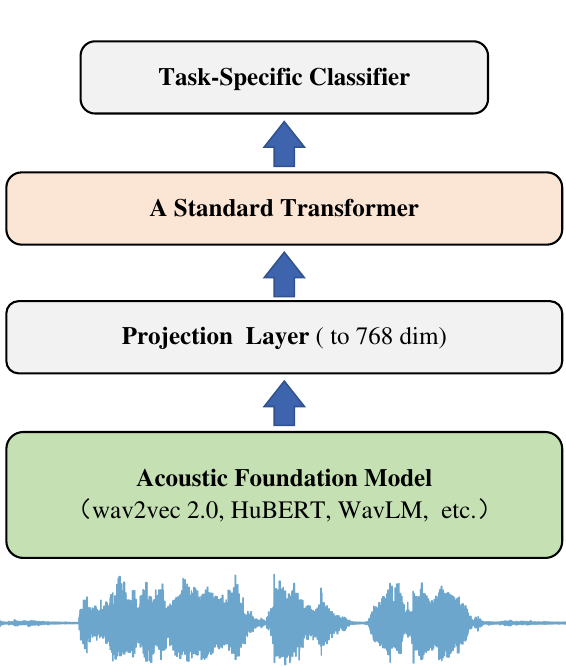}
  % \caption{The diagram illustrates the benchmark model employed by ParaLBench. Initially, it extracts deep features from the audio through an acoustic foundation model. Subsequently, these features are mapped to the working dimensions of the Transformer via a Map layer. Finally, using a task-specific classifier, probabilities for each category are obtained.}
  \caption{An evaluation diagram for ParaLBench. It first extracts deep representations from audio signals through an acoustic foundation model. These representations are then projected to the ones with a fixed dimension and sequentially fed into a standard Transformer. Finally, a task-specific classifier (\ie two dense layers) is appended for classification.}
  \label{fig:framework}
\end{figure}

The unified evaluation framework is illustrated in  \textcolor{black}{Figure}~\ref{fig:framework}, which comprises a pre-trained acoustic foundation model, a projection layer, a standard Transformer, and a task-specific classifier. Specifically, the acoustic foundation model is generally pre-trained via certain SSL algorithms, such as contrastive learning, denoising, and masked token prediction. The projection layer maps the representations from different acoustic foundation models into fixed-dimension ones to fit the standard input of the succeeded Transformer. The task-specific classifier consists of two dense layers. 

\reviewone{Compared to previous methods~\cite{yang2024large, yang2021superb}, we introduced a standard Transformer to enhance the ability to capture semantic information and contextual understanding. The self-attention mechanism of the Transformer effectively focuses on important semantic information within the input sequence, which improves the richness of feature representation. Furthermore, this structure is flexible and efficient, simplifying the model architecture while enhancing its adaptability to downstream tasks.}
% It uses a standard Transformer~\cite{vaswani2017attention} as the benchmark model to extract and learn high-level semantic representations of speech signals, facilitating more effective execution of downstream tasks. Subsequently, we feed the output of the Transformer into a classifier composed of two fully connected layers to obtain the final prediction results. 

Mathematically, this process can be described by the following formula:
\begin{equation}
% \text{Attention}(Q, K, V) = \text{softmax}\left(\frac{QK^T}{\sqrt{d_k}}\right)V
\mathbf{h} = f(\mathbf{x}),
\end{equation}
\begin{equation}
\tilde{\mathbf{h}} = \text{Transformer}(\mathbf{h} \mathbf{W}_{m}),
\end{equation}
\begin{equation}
p = \text{Classifier}(\tilde{\mathbf{h}}[0]),
\end{equation}
% where $ x_i $ represents the raw speech input of the i-th sample, $ f(\cdot) $ denotes the acoustic foundation model used, $ h_i\in \mathbb{R}^{L\times h} $ represents the extracted features. $W_{m}\in \mathbb{R}^{h\times d}$ is a learnable matrix that maps the feature to the Transformer work dimension d. $ \tilde{h_i}\in \mathbb{R}^{L\times d} $  denotes the output of the Transformer. $ p_{i}\in \mathbb{R}^{C} $ represents the probability distribution over the output classes by the classifier.
where $ \mathbf{x} $ represents the input speech features given a speech sample, $ f(\cdot) $ denotes the acoustic foundation model , $ \mathbf{h}\in \mathbb{R}^{L\times h} $ represents the extracted features, where $L$ denotes the feature sequence length and $h$ represents the dimension. $\mathbf{W}_{m}\in \mathbb{R}^{h\times d}$ is a learnable matrix that maps the feature to the Transformer input and output dimension $d$. $ \mathbf{\tilde{h}}\in \mathbb{R}^{L\times d} $  denotes the output of the Transformer. $ p\in \mathbb{R}^{C} $ represents the probability distribution over the output classes by the classifier. 

For handcraft features, the $ f(\cdot) $ denotes handcrafted Feature Extractor, and $ \mathbf{h}\in \mathbb{R}^{h}$ is obtained through $ f(\cdot) $. \editor{Then, we directly feed $ \mathbf{h}$ into the task-specific classifier, which consists of two dense layers, to obtain the final prediction:}

\begin{equation}
p = \text{Classifier}(\mathbf{h}).
\end{equation}
During training, we employ an end-to-end training approach. For classification tasks, we utilize the cross-entropy loss function $ \mathcal{L}_{CE} $ as the objective function, while for regression tasks, we employ the \textcolor{black}{mean absolute error (MAE)} loss $ \mathcal{L}_{MAE} $. $ \mathcal{L}_{CE} $ and $ \mathcal{L}_{MAE} $ are specifically defined as:
\begin{equation}
\mathcal{L}_{CE} = - \frac{1}{N} \sum_{i=1}^{N} \sum_{c=1}^{C} y_{i,c} \log(p_{i,c}),
\end{equation}
\begin{equation}
\mathcal{L}_{MAE} = \frac{1}{n} \sum_{i=1}^{n} \left| y_i - \hat{y}_i \right|
% \frac{1}{N} \sum_{i=1}^{N} (y_i - p_{i})^2,
\end{equation}
where $N$ is the number of samples, $C$ represents the number of classes, and $y_i$ represents the normalized true label of sample $i$.

\subsection{Acoustic Foundation Models}

% Different acoustic foundational models may have varying model architectures, training objectives, and training data. Consequently, diverse base models can capture distinct paralinguistic information during the training process. For instance, incorporating emotional data in the training process enhances the model's ability to learn potential emotional paralinguistic information. Additionally, different model architectures such as Convolutional Neural Networks (CNNs) excel at capturing local information and are thus more likely to grasp locale-specific paralanguage details. On the other hand, Transformer-based models due to the self-attention mechanism excel at capturing globally-dependent paralanguage information. This acquired paralinguistic knowledge holds significant potential for positively impacting computational tasks related to paralinguistics; hence selecting an appropriate phonological grounding model is particularly crucial for downstream tasks.

%  The 14 acoustic foundation models and 2 handcraft features we investigated in this study are comprehensively listed in Table~\ref{foundation model}. Our approach does not require the development of a novel underlying language model or dataset. Instead, we leverage a diverse range of resources from the open source community (e.g., Huggingface, Fairseq, Opensmile), all readily available for utilization and replication by subsequent researchers.

Nowadays, due to the development of SSL, various acoustic foundation models have emerged in the speech-processing society, with different neural network architectures, training objectives, and training data. 
Most of these models were extensively evaluated in ASR in previous years. Although these models are gradually and increasingly exploited in ComParal, it lacks a holistic and unified performance investigation of these foundation models in the application to ComParal. In the following, we will introduce two handcrafted feature sets and eight frequently used acoustic foundation models. As some of these models contain base, \textcolor{black}{medium}, % media
and large versions, we investigated 14 acoustic foundation models in total. The details of these models and feature sets can be found in Table~\ref{foundation model}. 

% Acoustic foundational models exhibit diverse architectures, training objectives, and training data, each capturing unique facets of paralinguistic information. For example, models trained with emotional data are better equipped to detect emotional paralinguistics. Architecturally, CNNs excellently capture local speech features, whereas Transformer-based models, with their self-attention mechanisms, adeptly process speech on a global scale. This variety necessitates careful selection of a model tailored to the specific paralinguistic task at hand. The 14 acoustic foundation models and 2 handcrafted feature sets evaluated in this study are detailed in Table~\ref{foundation model}, sourced from open community resources like Huggingface, Fairseq, and Opensmile.

\begin{table*}[t]
  \vspace{-0.3cm}
  \centering
  \setlength\tabcolsep{6pt}
  \caption{Details of the selected acoustic foundation models. LS: LibriSpeech; LL: LibriLight. }
  \label{foundation model}
  \begin{tabular}{cccrcc}
    \toprule
    Model &  Feature Types & Backbones & \# Params &  Training Datasets & Toolkits\\
    \midrule
    \midrule
     eGeMAPS&  Handcraft & - & - &  - & \textcolor{black}{openSMILE}\\
     ComParE-2016 &  Handcraft & - &  - & - & \textcolor{black}{openSMILE}\\
     \midrule
     wav2vec-large~\cite{schneider2019wav2vec}&  SSL&  CNNs& 32.54M & LS-960 ~\cite{panayotov2015librispeech} & Fairseq\\
     
     \multirow{2}*{emotion2vec-base~\cite{ma2023emotion2vec}}&  \multirow{2}*{SSL}&  \multirow{2}*{Transformer}& \multirow{2}*{93.79M} & IEMOCAP~\cite{busso2008iemocap}, MELD~\cite{poria2018meld}, MEAD~\cite{wang2020mead}, & \multirow{2}*{Fairseq}\\
     
     &  &   &  & CMU-MOSEI~\cite{zadeh2018multimodal}, MSP-Podcast~\cite{msp-podcast}  \\
     \midrule
     
     wav2vec2-base~\cite{baevski2020wav2vec}&  SSL&   Transformer& 95.04M & LS-960~\cite{panayotov2015librispeech} & Huggingface \\
     
     wav2vec2-large~\cite{baevski2020wav2vec}&  SSL&  Transformer& 317.38M & LS-960~\cite{panayotov2015librispeech} & Huggingface\\
     
     \multirow{2}*{wav2vec2-large-age-gender~\cite{burkhardt2023speech}}&  \multirow{2}*{SSL}&  \multirow{2}*{Transformer}& \multirow{2}*{317.38M}  &  aGender~\cite{burkhardt2010database}, Common Voice~\cite{ardila2019common},  & \multirow{2}*{Huggingface}\\
      &  &   &  & Timit~\cite{garofolo1993timit}, Voxceleb 2~\cite{nagrani2020voxceleb}  \\
     
     wav2vec2-large-xlsr-53~\cite{conneau2020unsupervised}&  SSL&  Transformer& 317.38M & 
Common Voice~\cite{ardila2019common}, BABEL, MLS~\cite{pratap2020mls}  & Huggingface \\

     WavLM-base~\cite{chen2022wavlm}&SSL&  Transformer& 94.70M & LS-960~\cite{panayotov2015librispeech} & Huggingface\\
     
     WavLM-large~\cite{chen2022wavlm}&  SSL&  Transformer& 316.62M & LL-60k~\cite{kahn2020libri}, VoxPopuli~\cite{wang2021voxpopuli}, GigaSpeech~\cite{chen2021gigaspeech} & Huggingface\\
     
     Whisper-base~\cite{radford2022whisper}&  SSL&  Transformer& 71.83M & MTD-680k~\cite{radford2022whisper} & Huggingface\\  % Multitask training data (680k hours)
     
     Whisper-large~\cite{radford2022whisper}&  SSL&  Transformer&  1541.38M & MTD-680k~\cite{radford2022whisper}& Huggingface\\
     
     HuBERT-base~\cite{hsu2021hubert}&  SSL&  Transformer& 94.68M & LS-960~\cite{panayotov2015librispeech}& Huggingface\\
     
     HuBERT-large~\cite{hsu2021hubert}&  SSL&  Transformer& 316.61M & LL-60k~\cite{kahn2020libri}& Huggingface \\
     
     \multirow{2}*{CLAP-HTSAT~\cite{wu2023large}}&  \multirow{2}*{SSL}&  \multirow{2}*{Transformer}& \multirow{2}*{153.49M} & AudioCaps~\cite{drossos2020clotho}, Clotho~\cite{kim2019audiocaps}, & \multirow{2}*{Huggingface}\\
     &  &   &  & LAION-Audio-630K~\cite{wu2023large}, Audioset~\cite{gemmeke2017audio} \\
     
     data2vec-audio-base~\cite{baevski2022data2vec}&  SSL&  Transformer& 93.16M & LS-960 ~\cite{panayotov2015librispeech} & Huggingface\\
    \bottomrule
  \end{tabular}
\end{table*}

\subsubsection{Handcraft features}
Early ComParal research typically involved designing handcrafted features to accommodate various tasks.
Given the widespread utility of \editor{eGeMAPS (Extended GeMAPS)}~\cite{trigeorgis2016adieu} and ComParE-2016 (Computational Paralinguistics Challenge 2016)~\cite{powell2022deep,edwards2020multiscale,10446795} in the field of speech processing, we contrasted them with language-based models in this study. eGeMAPS captures emotion-relevant acoustic properties, while ComParE-2016 offers a broader suite for tasks like emotion and gender recognition. These features, accessible via the \editor{openSMILE} toolkit, provide a rich tapestry of speech-based information.

\subsubsection{wav2vec}
wav2vec~\cite{schneider2019wav2vec} is an acoustic foundation model entirely composed of convolutional neural networks, which takes raw audio as input and is trained on a large amount of unlabeled audio data. Its training objective is to use a contrastive loss function to distinguish between real future audio samples and negative samples, thus obtaining a universal representation suitable for input into speech recognition systems. This foundational model is readily available on the open-source community Fairseq.

\subsubsection{wav2vec 2.0}
wav2vec 2.0~\cite{baevski2020wav2vec} represents an improved version of wav2vec. Its main improvements include: 1. Replacing the convolutional neural network with a Transformer, enhancing the model's ability to represent context. 2. Discretizing the output of the feature encoder z into a finite set of speech representations through quantization. 3. It masks the latent representation of the original waveform and solves contrastive learning tasks through quantized speech representations. Currently, wav2vec 2.0 has several derivative versions, such as wav2vec2-large-age-gender~\cite{burkhardt2023speech} fine-tuned on gender and age datasets, and wav2vec2-large-xlsr-53~\cite{conneau2020unsupervised} fine-tuned on multilingual datasets. These foundational models are all available for download on the open-source community Huggingface.

\subsubsection{emotion2vec}
emotion2vec~\cite{ma2023emotion2vec} serves as a universal speech emotion representation model. This model is pre-trained on multiple open-source unlabeled emotion datasets through self-supervised online distillation, combining utterance-level loss and frame-level loss during pre-training. As a result, emotion2vec demonstrates consistent improvements across multiple speech emotion recognition datasets. This foundational model is conveniently accessible on the open-source community Fairseq.

\subsubsection{HuBERT}
HuBERT~\cite{hsu2021hubert} utilizes an offline clustering step to generate pseudo-labels for pre-training, following a methodology akin to BERT.
Specifically, the HuBERT model employs masked continuous speech features to predict predetermined cluster labels.
This predictive loss is exclusively applied within the masked regions, compelling the model to acquire robust high-level representations and enabling accurate inference of the masked targets.
Inspired by the DeepCluster~\cite{caron2018deep} method in self-supervised visual learning, HuBERT leverages the masked prediction loss over speech sequences to encapsulate their sequential structure.
Presently, this foundational model is available for download on the open-source community Huggingface.

\subsubsection{WavLM} 
WavLM~\cite{chen2022wavlm} introduces a masking-based framework for speech denoising and prediction.
Within this framework, certain inputs simulate noisy or overlapping speech with masks, aiming to predict pseudo-labels for the original speech within the masked regions, akin to HuBERT.
The framework integrates masked speech prediction and denoising during pre-training. Consequently, the WavLM model not only learns ASR information through masked speech prediction but also acquires knowledge of non-ASR tasks via speech denoising modeling. 
Moreover, WavLM optimizes the model structures and training data of HuBERT and Wav2vec 2.0.
Specifically, it incorporates gated relative position biases (grep) into the Transformer structure as the backbone, enhancing the model's ASR performance while maintaining nearly identical parameter count and training speed. Furthermore, to bolster model robustness and mitigate data mismatch issues, the unlabeled pre-training data is expanded to encompass 94k hours of public audio.

\subsubsection{CLAP}
CLAP~\cite{wu2023large}, similar to CLIP~\cite{radford2021learning},  employs the contrastive learning paradigm to train models on large-scale noisy data collected from the internet. The training process of CLAP involves encoding raw speech and text into high-level representations using separate speech and text encoders. Genuine speech-text pairs are treated as positive samples, while the rest are considered negative, and contrastive loss is computed accordingly. 
CLAP uses the Transformer as the core architecture for the speech encoder and introduces a feature fusion scheme to handle variable-length audio inputs.
In this study, we primarily extract speech features through CLAP’s audio encoder.

\subsubsection{Whisper}
Whisper~\cite{radford2022whisper} represents a significant advancement in audio modeling, released by OpenAI in 2022. It encompasses functionalities such as multilingual ASR, speech translation, and language identification.
The Whisper model has showcased exceptional recognition performance, even with large amounts of weakly labeled data, without the necessity for complex models and tuning methods, particularly excelling in robustness and generalization.
Diverging from pre-trained speech models trained using unsupervised methods like Wav2vec 2.0, Whisper employs weakly supervised training, enabling direct multitask learning without the need for task-specific fine-tuning.

\subsubsection{Data2vec}
Self-supervised learning has made progress in computer vision, natural language processing, and speech processing, but algorithms designed for a single modality obscure cross-modal learning mechanisms. Therefore, Data2vec~\cite{baevski2022data2vec} proposes enhancing the computational efficiency of self-supervised learning through contextualized target prediction, efficient data encoding, and fast convolutional decoders. 
It employs a unified learning objective but trains separate models for different modalities using Transformer architecture and different feature encoders.
It utilizes a teacher model to create latent contextualized representations, which are regressed by a student model using masked examples. Contextualized targets capture sample information, enriching learning tasks and speeding up learning. Experimental results show efficiency improvements of 2 to 16 times with similar accuracy in tasks like image classification, speech recognition, and natural language understanding.
% Experimental results demonstrate efficiency improvements ranging from 2 to 16 times with comparable accuracy in tasks such as image classification, speech recognition, and natural language understanding.

\subsection{The ParaLBench Tasks}
We categorize the tasks in ParaLBench using the time axis taxonomy method of computational psycholinguistics as outlined in~\cite{schuller2013paralinguistics, schuller2013computational}. 
\reviewthree{Based on the period of the impact of the paralinguistic information on the voice articulation, we group the paralinguistic information into long-, median-, and short-term groups. For example, the age and gender often alter the speech characteristics through years; while emotion often can transform the articulation behavior instantaneously, from few seconds to minutes.}
Traits requiring extended periods for determination are classified as ``long-term'', encompassing biological trait primitives, cultural traits, personality traits, and aspects of the speaker's personal style.
States ascertainable in the short term are designated as ``short-term'', encompassing emotions and associated states such as stress, confidence, and uncertainty.
``Medium-term'' traits and states typically involve self-induced characteristics such as sleepiness, intoxication (\eg alcohol poisoning), and health states, or may entail discrepant speech patterns such as sarcasm and lying.
For detailed task taxonomies and corresponding datasets, please refer to Table~\ref{tab:dataset_taxonomy}.

\begin{table}[!t]
  \setlength\tabcolsep{11.5pt}
  \centering
  % \caption{The paralinguistic task taxonomy.}
  \caption{A taxonomy of the selected paralinguistic tasks for ParaLBench.}
  \label{tab:dataset_taxonomy}
  \begin{tabular}{c| ccc}
    \toprule
    \textbf{Taxonomy} & \textbf{Tasks} & \textbf{Datasets} & \textbf{\editor{\# Classes}}\\
    \midrule
    \midrule
    \multirow{8}*{Short-term} &  Emotion & MELD & 7\\
    &  Emotion & MSP-Podcast & 5 \\
    &  \reviewtwo{Emotion} & \reviewtwo{IEMOCAP} & \reviewtwo{4} \\
    & \reviewtwo{Arousal} & \reviewtwo{IEMOCAP} & \reviewtwo{1*}\\
    &  \reviewtwo{Valence} & \reviewtwo{IEMOCAP} & \reviewtwo{1*}\\
    &  \reviewtwo{Dominance} & \reviewtwo{IEMOCAP} & \reviewtwo{1*}\\
    &  Sentiment & MELD & 3\\
    &  Sentiment & CMU-MOSI & 2\\
    \midrule
    % \multirow{4}*{Medium-term} & Speaking Speed  & PromptSpeech & 3\\
    \multirow{4}*{Medium-term} & Sarcasm &  MUStARD & 2 \\
    & Inﬂuenza &  FluSense & 9 \\
    & Stutter & SEP-28K & 2 \\
    & \reviewtwo{Depression} &\reviewtwo{DAIC-WOZ} & \reviewtwo{2}\\
    
    \midrule
    \multirow{4}*{Long-term} &  Gender& MSP-Podcast & 2\\
    % & Gender & PromptSpeech & 2\\
    & Age & VCTK & 1*\\
    & Accent & VCTK & \textcolor{black}{11}\\
    % & Region & VCTK \\
    & Dialect & TIMIT & 8\\
    \bottomrule
  \end{tabular}

  \begin{tablenotes}
  \item[1] * Regression only
  \end{tablenotes}
\end{table}

\subsubsection{Emotion}
SER~\cite{chen2023speechformer++, 8462579} refers to automatically inferring the speaker's emotional state by analyzing both the acoustic features and linguistic content of the speech signal. 
This task holds diverse applications in affective computing, human-computer interaction, customer service, and various other domains.
%Effective emotion recognition can significantly enhance systems' understanding and responsiveness to user emotions, thereby improving user experience and interaction effectiveness.  这里删掉了
For this task, \reviewtwo{we will utilize the  MELD, IEMOCAP, and MSP-Podcast datasets.}

% Speech Emotion Recognition~\cite{chen2023speechformer++, 8462579} involves inferring a speaker's emotional state by analyzing the acoustic features and linguistic content. Applied in affective computing and human-computer interaction, this task improves system response to human emotions. The MELD and MSP-Podcast datasets are employed here.

\subsubsection{Emotion Dimensions}
\reviewtwo{Arousal, Valence, and Dominance are key dimensions for describing emotions. Arousal measures the intensity or energy level of an emotion, ranging from calm to excited. Valence indicates the positive or negative nature of the emotion, from pleasant to unpleasant. Dominance reflects the sense of control or power one feels in a situation, varying from feeling submissive to dominant. Together, these dimensions provide a comprehensive framework for understanding and analyzing emotional states. For these tasks, we will utilize the IEMOCAP dataset.}

\subsubsection{Sentiment}
% Sentiment analysis~\cite{10447044} refers to the process of automatically identifying and categorizing the emotional tendencies in speech using machine learning techniques.Its primary objective is to determine the emotional polarity of speech, whether positive or negative.Sentiment analysis tasks find extensive applications in fields such as social media monitoring, sentiment analysis, and product review analysis. For this task, we will utilize the MELD dataset and the CMU-MOSI dataset.
\reviewtwo{Sentiment analysis~\cite{10447044} involves automatically identifying and categorizing emotional tendencies in speech using machine learning techniques. Its primary objective is to determine the emotional polarity of speech as positive or negative. This task has applications in social media monitoring and product review analysis. We will utilize the MELD and CMU-MOSI datasets for this task.}

\subsubsection{Sarcasm}
% The Sarcasm task~\cite{joshi2017automatic} aims to detect and identify sarcastic speech in audio. Sarcasm is a mode of expression often used to convey the opposite or contradictory viewpoint of the speaker's true intention. Sarcasm tasks hold significant importance in social media analysis, sentiment monitoring, and other fields. By recognizing sarcastic speech, a better understanding of the true meaning and intentions of speakers within audio data can be achieved. For this task, we will utilize the MUStARD dataset.
\reviewtwo{The Sarcasm Detection Task~\cite{joshi2017automatic} focuses on identifying sarcasm in audio, a form of expression conveying the speaker’s opposite intention, critical in social media and sentiment analysis.}

\subsubsection{Influenza}
Influenza is a common respiratory disease caused by influenza viruses that lead to large outbreaks of infection and disease worldwide every year. The purpose of Influenza analysis~\cite{chang2022covnet} is to analyze and predict influenza-related information through speech processing and machine learning technology.

\subsubsection{Stutter}
% Stutter Analysis: Stuttering is a speech disorder characterized by disruptions and interruptions during speech, typically involving repetition of syllables, words, and elongated sounds. The Stutter analysis task aims to analyze and understand the features, patterns, and influencing factors of stuttering. By analyzing the features and patterns of stuttering~\cite{kourkounakis2021fluentnet}, a better understanding of its underlying mechanisms and influencing factors can be gained, providing more effective assistance and support for stuttering patients.
\reviewtwo{Stuttering is a speech disorder marked by disruptions during speech, often involving repetition of syllables, words, and elongated sounds. The stutter analysis task aims to understand the features and patterns of stuttering~\cite{kourkounakis2021fluentnet}. This analysis can reveal underlying mechanisms and influencing factors, ultimately providing more effective support for individuals who stutter.}

\subsubsection{Depression}
%\reviewtwo{The task of depression detection~\cite{sardari2022audio, fan2024transformer} aims to identify the presence of depressive symptoms by analyzing an individual's emotional, behavioral, and cognitive data. The goal is through machine learning technology to accurately detect signs of depression to enable timely intervention and personalized treatment, ultimately improving the individual's mental health and quality of life. For this task, we will utilize the DAIC-WOZ dataset.}
\reviewtwo{Depression detection~\cite{sardari2022audio, fan2024transformer} involves identifying depressive symptoms by analyzing an individual's emotional, behavioral, and cognitive data. The aim is to leverage machine learning to accurately detect signs of depression, enabling timely intervention and personalized treatment. For this task, we will utilize the DAIC-WOZ dataset.}

\subsubsection{Gender}
The Gender task~\cite{radha2024automatic, alamsyah2020speech} aims to identify and analyze the gender characteristics of speakers in speech. This task has wide applications in fields such as speech recognition, gender identification technology, and social media analysis, enabling us to better understand and utilize gender information in speech data. For this task, we will utilize the MSP-Podcast dataset.

\subsubsection{Age}
The Age task~\cite{radha2024automatic} aims to analyze and identify the age characteristics of speakers through speech signals. This task has broad applications in fields such as speech recognition, age identification technology, and targeted advertising, enabling us to better understand and utilize age information in speech data. For this task, we will utilize the VCTK dataset.

\subsubsection{Accent}
% Accent refers to the individual characteristics of pronunciation and speech rhythm exhibited by speakers, typically influenced by factors such as their native language, region, and cultural background. The Accent task~\cite{mikhailava2022language} aims to identify and analyze the accent features of speakers in speech using artificial intelligence technology. This task holds broad applications in fields such as speech recognition, linguistics research, and cross-cultural communication. For this task, we will utilize the VCTK dataset.
\reviewtwo{Accent refers to the unique pronunciation and speech rhythm of speakers, influenced by their native language, region, and culture. The Accent task~\cite{mikhailava2022language} aims to identify and analyze these features using artificial intelligence. This task has applications in speech recognition, linguistics research, and cross-cultural communication. For this task, we will utilize the VCTK dataset.}

\subsubsection{Dialect}
Dialect refers to language variants used in specific regions or communities, encompassing differences in vocabulary, grammar, pronunciation, and other linguistic aspects. The Dialect analysis task~\cite{lin2020transformer} holds significant importance in fields such as speech recognition, cultural preservation, and sociolinguistic research. For this task, we will utilize the TIMIT dataset. 

By leveraging these categorized tasks and datasets, ParaLBench strives to enhance ComParal research, facilitating a comprehensive understanding of human speech patterns and behaviors.

\subsection{The Unified Evaluation Metrics}
Considering the diverse evaluation metrics used in prior approaches\cite{pramanick2022multimodal, chen2023speechformer++, lea2021sep}, we take unified evaluation metrics in ParaLBench to facilitate the comparison of various acoustic foundation models across different computational paralinguistic tasks. Simultaneously, we encourage future researchers to adhere to the same settings and evaluation criteria to ensure fair comparisons of their algorithms.

For classification tasks, we employ \editor{Weighted Accuracy (WA), Unweighted Accuracy (UA), and Weighted F1 Score (WF1)} as evaluation metrics. The aforementioned metrics can be described as follows:
\begin{equation}
% WA = \frac{{\sum_{i=1}^{N} w_i \cdot \mathbb{I}(y_i = \hat{y}_i)}}{{\sum_{i=1}^{N} w_i}}
\text{WA} = \frac{1}{\sum_{c=1}^{C} N_c} \sum_{c=1}^{C} N_c \times Acc(c),
\end{equation}

\begin{equation}
% UA = \frac{{\sum_{i=1}^{N} \mathbb{I}(y_i = \hat{y}_i)}}{N}
\text{UA} = \frac{1}{C} \sum_{c=1}^{C} Acc(c),
\end{equation}

\begin{equation}
%WF1 = \frac{{2 \cdot \sum_{i=1}^{N} w_i \cdot \text{precision}_i \cdot \text{recall}_i}}{{\sum_{i=1}^{N} w_i \cdot (\text{precision}_i + \text{recall}_i)}}
\text{WF1} = \frac{1}{\sum_{c=1}^{C} N_c} \sum_{c=1}^{C} N_c \times F1(c),
\end{equation}
where $N_c$ represents the number of samples in class $c$, $Acc(c)$ and $F1(c)$ respectively denote the classification accuracy and F1 score of class $S_c$.

The \textcolor{black}{Mean Absolute Error (MAE)} is used as the evaluation metric for regression tasks and the formula can be described as follows:  
\begin{equation}
% MSE = \frac{1}{N} \sum_{i=1}^{N} (y_i - \hat{y}_i)^2,
\text{MAE} = \frac{1}{n} \sum_{i=1}^{n} \left| y_i - \hat{y}_i \right|
\end{equation}
where $N$ represents the number of samples, $y_i$ denotes the true labels, and $\hat{y}_i$ represents the predicted values.

Higher values of WA, UA, and WF1 indicate better performance, whereas lower values of MAE denote superior performance.

\section{Experimental Setup}
\label{sec:experiment}
In this section, we introduce the datasets used in ParaLBench, as shown by Table~\ref{tab:dataset}, as well as the method of dividing the datasets. Then we describe the implementation details.

\subsection{ParaLBench Corpus}

\begin{table*}[htbp]
  \vspace{-0.2cm}
  \setlength\tabcolsep{14pt}
  \centering
  \caption{Details of the selected datasets for the ParaLBench.}
  \label{tab:dataset}
  \begin{tabular}{c| crr|rrr}
    \toprule
    \textbf{Datasets} & \textbf{Paralinguistic Tasks} & \textbf{Duration}& \textbf{\# Total} & \textbf{\# Training} & \textbf{\# Validation} & \textbf{\# Test}\\
    \midrule
    \midrule
    MELD~\cite{poria2018meld}&  Emotion, Sentiment & 12.18 hr  & 13,708  &  9,986 & 1,108  & 2,609   \\
    MSP-Podcast~\cite{msp-podcast}& Emotion, Gender  &  26.15 hr & 90,978 & 63,076  &  10,999  & 16,903   \\
    CMU-MOSI~\cite{zadeh2018multimodal}& Sentiment  &  2.62 hr &  2,199 & 1,284  &  229 &  686  \\
    \reviewtwo{IEMOCAP~\cite{busso2008iemocap}} & \reviewtwo{Emotion} & \reviewtwo{6.99 hr} & \reviewtwo{5,531} & \reviewtwo{4,290} & \reviewtwo{-} & \reviewtwo{1,241} \\
    \reviewtwo{IEMOCAP~\cite{busso2008iemocap}} & \reviewtwo{Arousal, Valence, Dominance} & \reviewtwo{12.44 hr} &\reviewtwo{10,039} & \reviewtwo{7,869} & \reviewtwo{-} & \reviewtwo{2,170} \\
    
    % PromptSpeech~\cite{guo2023prompttts} &  Speaking Speed, Gender & 59.52 hr  & 27881  &  26576 & - &  1305 \\
    % CSTR VCTK~\cite{veaux2016superseded}&  Age, Accent, Region & 4.37 hr  &  44,070 &  39,663 &  - & 4,407 \\
    MUStARD~\cite{mustard}&  Sarcasm &  2.67 hr &  690  &  334 & -  &  356  \\
    FluSense~\cite{10.1145/3381014}&  Influenza &  14.10 hr &  11,602  &  9,281  &  - & 2,321  \\
    SEP-28k~\cite{lea2021sep}& Stutter  &  23.26 hr &  27,922 & 24,922  &  2,000 & 1,000    \\

    \reviewtwo{DAIC-WOZ~\cite{gratch2014distress}}& \reviewtwo{Depression} & \reviewtwo{50.21 hr} & \reviewtwo{189} & \reviewtwo{107} & \reviewtwo{35} & \reviewtwo{47} \\
    
    CSTR VCTK~\cite{veaux2016superseded}&  Age, Accent & \textcolor{black}{43.88} hr  &  44,070 &  39,663 &  - & 4,407 \\
    TIMIT~\cite{garofolo1993timit}&  Dialect & 5.38 hr &  6,300  & 4,620  & -  & 1,680   \\
    \bottomrule
  \end{tabular}
\end{table*}

\reviewtwo{In this study, the specific datasets chosen for evaluation tasks were primarily based on their representativeness, applicability, and ease of access. This ensures that the benchmarks established are not only relevant but also practical for future research in the field.}

\subsubsection{MELD}
% The Multimodal EmotionLines Dataset~\cite{poria2018meld} [71] is an extension of the EmotionLines dataset, enriched with audio and visual modalities. It comprises 13,708 dialogue segments extracted from the TV series ``Friends'', involving multiple speakers. The dataset is annotated with seven emotional states: Anger, Disgust, Sadness, Joy, Neutral, Surprise, and Fear. Additionally, MELD provides sentiment polarity annotations (positive, negative, and neutral) for each utterance. Similarly, we conduct experiments following the official dataset partitioning and report results on the test set.
\reviewtwo{The Multimodal EmotionLines Dataset~\cite{poria2018meld} [71] is an extension of the EmotionLines dataset. It comprises 13,708 dialogue segments. The dataset is annotated with seven emotional states: Anger, Disgust, Sadness, Joy, Neutral, Surprise, and Fear. Additionally, MELD provides sentiment polarity annotations (positive, negative, and neutral) for each utterance. Similarly, we conduct experiments following the official dataset partitioning and report results on the test set.}

\subsubsection{MSP-Podcast}
% MSP-Podcast~\cite{msp-podcast} stands out as one of the largest speech-emotion corpora available in the research community. It encompasses speech segments extracted from podcast recordings, perceptually annotated through crowdsourcing, and continually updated. For this study, we employ version v1.10 of the corpus, comprising 104,267 speech segments. The dataset is officially partitioned into Train, Development, Test1, and Test2 sets. Test results on Test1 will be disclosed following previously established methodologies~\cite{tompkins2023multi}. Annotations within the dataset encompass attribute-based descriptors (activation, dominance, and valence) alongside categorical labels (anger, happiness, sadness, disgust, surprised, fear, contempt, neutral, and others).
\reviewtwo{MSP-Podcast~\cite{msp-podcast}, one of the largest speech-emotion corpora, contains speech segments from podcast recordings, annotated via crowdsourcing and regularly updated.  For this study, we employ version v1.10 of the corpus, comprising 104,267 speech segments. The dataset is officially partitioned into Train, Development, Test1, and Test2 sets. Test results on Test1 will be disclosed following previously established methodologies~\cite{tompkins2023multi}. Annotations within the dataset encompass categorical labels (anger, happiness, sadness, disgust, surprised, fear, contempt, neutral, and others).}

\subsubsection{CMU-MOSI}
% The Multimodal Corpus of Sentiment Intensity (CMU-MOSI)~\cite{zadeh2018multimodal} dataset consists of 2,199 opinion video clips, each annotated with sentiment scores ranging from -3 to 3. It encompasses comprehensive annotations for subjectivity and sentiment intensity, along with visual features annotated per-frame and per-opinion, and audio features annotated per-millisecond. We follow the official dataset partitioning to train models using the training set, tune hyperparameters using the validation set, and ultimately select the best-performing model on the validation set for evaluation on the test set, reporting the results thereafter.
\reviewtwo{The Multimodal Corpus of Sentiment Intensity (CMU-MOSI)~\cite{zadeh2018multimodal} dataset consists of 2,199 opinion video clips, each annotated with sentiment labels (Positive, Neutral, Negative).  We follow the official dataset partitioning to train models using the training set, tune hyperparameters using the validation set, and ultimately select the best-performing model on the validation set for evaluation on the test set, reporting the results thereafter.}

\subsubsection{IEMOCAP}
\reviewtwo{The Interactive Emotional Dyadic Motion Capture (IEMOCAP)~\cite{busso2008iemocap} dataset is a multimodal resource for emotion recognition. It consists of a series of session dialogue scenarios capturing interactions among multiple participants in various emotional states. Additionally, the dataset provides information on emotional dimensions e.g. Arousal, Valence, and Dominance. In this paper, we use the first four sessions as the training set and the last session as the test set to guarantee speaker independence.
In emotion prediction, we utilize four types of data: angry, neutral, happy (\& excited), and sad.}

\subsubsection{MUStARD}
% The MUStARD~\cite{mustard} dataset is a multimodal video corpus designed for automatic sarcasm detection research, sourced from popular TV shows like ``Friends'', ``Golden Girls'', ``The Big Bang Theory'', and ``Sarcasmaholics Anonymous''. It comprises 690 audiovisual utterances labeled with sarcasm. Each utterance comes with contextual information providing additional details about the scene. \reviewtwo{We utilize the speaker-independent evaluation approach proposed in~\cite{mustard}}, using data from ``Golden Girls'', ``The Big Bang Theory'', and ``Sarcasmaholics Anonymous'' for training and data from ``Friends'' for testing.
\reviewtwo{The MUStARD~\cite{mustard} dataset is a multimodal video corpus designed for automatic sarcasm detection research, sourced from popular TV shows like ``Friends'', ``Golden Girls'', ``The Big Bang Theory'', and ``Sarcasmaholics Anonymous''. It comprises 690 audiovisual utterances labeled with sarcasm. \reviewtwo{We utilize the speaker-independent evaluation approach proposed in~\cite{mustard}}, using data from ``Golden Girls'', ``The Big Bang Theory'', and ``Sarcasmaholics Anonymous'' for training and data from ``Friends'' for testing.}

\subsubsection{FluSense}
% The FluSense~\cite{10.1145/3381014} dataset is based on the Google Audioset~\cite{gemmeke2017audio}, which contains 10-second audio clips from YouTube covering coughs, sneezes, speech, and various human activities and background noises. Since the timing of coughs and speech is not annotated, two annotators accurately labeled 45,550 seconds of the Audioset dataset. Following the method in~\cite{chang2022covnet}, we divided the dataset into training and testing sets in a ratio of 0.8/0.2 and excluded categories with fewer than 100 samples. Therefore, we used audio samples from nine categories: other, breathe, cough, gasp, silence, sneeze, sniffle, speech, throat-clearing.
\reviewtwo{The FluSense~\cite{10.1145/3381014} dataset is based on the Google Audioset~\cite{gemmeke2017audio}, which contains 10-second audio clips from YouTube covering coughs, sneezes, speech, and various human activities and background noises. Following the method in~\cite{chang2022covnet}, we divided the dataset into training and testing sets in a ratio of 0.8/0.2 and excluded categories with fewer than 100 samples. Therefore, we used audio samples from nine categories: other, breathe, cough, gasp, silence, sneeze, sniffle, speech, throat-clearing.}

\subsubsection{SEP-28k}
% SEP-28k~\cite{lea2021sep} is a dataset comprising over 28,000 clips labeled with five event types including blocks, prolongations, sound repetitions, word repetitions, and interjections. The audio is sourced from public podcasts predominantly featuring individuals who stutter interviewing others who stutter. In this dataset, we conduct a binary classification task, distinguishing between stuttering and non-stuttering instances. The dataset is partitioned according to the approach outlined in~\cite{lea2021sep}, with 25k samples allocated for training, 2k for validation, and 1k for testing.

\reviewtwo{SEP-28k~\cite{lea2021sep} is a dataset comprising over 28,000 clips labeled. The audio is sourced from public podcasts predominantly featuring individuals who stutter interviewing others who stutter. In this dataset, we conduct a binary classification task, distinguishing between stuttering and non-stuttering instances. The dataset is partitioned according to the approach outlined in~\cite{lea2021sep}, with 25k samples allocated for training, 2k for validation, and 1k for testing.}

\subsubsection{DAIC-WOZ}
\reviewtwo{The DAIC-WOZ (Depression and Anxiety Interview Corpus - Wizard of Oz) dataset is a resource for studying depression and anxiety. It includes dialogues with virtual agents, capturing video, audio, and text data to simulate mental health assessment scenarios. The dataset assists researchers in evaluating emotional states and mental health, particularly for automated detection of depression and anxiety.}

\subsubsection{CSTR VCTK}
% CSTR VCTK~\cite{veaux2016superseded} corpus contains speech data from 110 English speakers with different accents. The dataset comprises a total of 44,070 utterances, with each speaker reading approximately 400 sentences selected from inspired passages of newspapers, Rainbow passages, and speech accent archives. The newspaper texts are sourced from The Glasgow Herald and are used with permission from the Herald and Times Group. Since CSTR VCTK does not provide an official partitioning of the dataset, we employ previous methods~\cite{lian2022robust}, randomly selecting 90\% of speakers for training and the remaining 10\% for testing.
\reviewtwo{CSTR VCTK~\cite{veaux2016superseded} corpus contains speech data from 110 English speakers with different accents. The dataset comprises a total of 44,070 utterances, with each speaker reading approximately 400 sentences selected from inspired passages of newspapers, Rainbow passages, and speech accent archives. Since CSTR VCTK does not provide an official dataset partitioning, we employ previous methods~\cite{lian2022robust, 9746698}, randomly split 90\% of data for training and the remaining 10\% for testing.}

\subsubsection{TIMIT}
The TIMIT~\cite{garofolo1993timit} Acoustic-Phonetic Continuous Speech Corpus is a standard dataset used to evaluate the performance of automatic speech recognition systems. It comprises recordings of 10 phonetically rich sentences read by 630 speakers representing 8 different American English dialects. In our study, we conduct dialect classification experiments, following the dataset's official partitioning, training our models on the training set, and reporting results on the test set.

\subsection{Implementation Details}
To fairly compare the performance of different acoustic foundation models, we strive to ensure that the models are trained under identical settings as much as possible. During training, we set the working dimension $d$ of the Transformer in the baseline model to 768 and employed the AdamW optimizer to optimize model parameters. For deep features, we set the initial learning rate to \editor{$5\times 10^{-4}$}, while for handcrafted features, we observed that a larger learning rate could lead to significant fluctuations in performance. Hence, we opt for \editor{$5\times 10^{-5}$} as the initial learning rate. We utilize a polynomial decay strategy to dynamically adjust the learning rate throughout the entire training process. We set the maximum number of epochs to 60 and simultaneously set the weight decay to $10^{-2}$. \editor{To prevent overfitting, we introduced 0.1 Dropout rate in the standard Transformer layers and set it to 0.5 Dropout rate in the classification layer.}

\section{Results and Discussion}
\label{sec:result}
% In this section, we established a large-scale computational paralinguistic benchmark for acoustic foundation models and report results in Table~\ref{tab: short-term}, ~\ref{tab: medium-term}, and ~\ref{tab: long-term}. This benchmark includes 9 datasets, 11 paralinguistic tasks, and 14 acoustic foundational models. We hope this benchmark can guide future researchers in feature selection and point the way towards developing speech models with stronger generalization capabilities.

% In this section, we introduce a comprehensive computational paralinguistic benchmark tailored to evaluate acoustic foundational models, and we present our findings in Tables~\ref{tab: short-term},~\ref{tab: medium-term}, and ~\ref{tab: long-term}. Encompassing 8 datasets, 9 paralinguistic tasks, and 14 models, this benchmark aims to aid future research in feature selection and promote the refinement of speech models with robust generalization capabilities.

In this section, we present the results from large-scale experiments with thirteen paralinguistic tasks with 14 acoustic foundation models and two manually engineered feature sets, and provide detailed result analysis and discussion.

% \subsection{The ParaLBench}
% \label{sec:Benchmark}

% 数据相同的重新跑一遍。
\begin{table*}[htbp] % 使用table*环境创建一个跨越两栏的表格
    \setlength\tabcolsep{8.1pt}
    \centering
    % \caption{The Short-term task result for the ParaLBench}
    \caption{Results of the selected \textit{short-term} paralinguistic tasks (\ie emotion [emo.] and sentiment [sent.]) in ParaLBench over two handcrafted feature sets and 14 acoustic foundation models.}
    \label{tab: short-term}
    \resizebox{\textwidth}{!}{%
    \begin{tabular}{c|cccccccccccc}
    \toprule
    \multirow{2}*{Foundation models}&  \multicolumn{3}{c}{MELD (Emo., 7-cls)} &  \multicolumn{3}{c}{MSP-Podacst (Emo., 5-cls)} &  \multicolumn{3}{c}{MELD (Sent., 3-cls)} &  \multicolumn{3}{c}{CMU-MOSI (Sent., 2-cls)}\\
     &  WA&  UA&  WF1&   WA&  UA&  WF1&   WA&  UA&  WF1&   WA&  UA&  WF1 \\
    \midrule
    \midrule
     eGeMAPS & .481 & .143 & .313 & \editor{.502} & \editor{.218} & \editor{.379} & \editor{.491} & \editor{.347} & \editor{.351} & .531 & .527 & .535\\
     ComParE-2016& .481 & .143 & .312 & .475 & .200 & .307 & .481 & .333 & .312 & .464 & .537 &.387\\
     \midrule
     CLAP-HTSAT& .482 & .146 & .319 & .514 & .243 & .455 & .484 & .356 & .386 & .528 & .534 &.532\\
     data2vec-audio-base& .498 & .176 & .367 & .559 & .294 & .507 & .499 & .380 & .420 & .641 & .615 &.635\\
     emotion2vec-base& \textcolor{red!100}{$.531$} & .247 & \textcolor{red!70}{$.465$} & .586 & .342 & .546 & \textcolor{red!70}{$.561$} & \textcolor{red!100}{$.503$} & \textcolor{red!100}{$.547$} & \textcolor{red!100}{$.711$} & \textcolor{red!70}{$.676$} & \textcolor{red!70}{$.699$}\\
     HuBERT-base& .481 & .216 & .420 & .579 & .347 & .538 & .519 & .434 & .483 & \textcolor{red!45}{$.676$} & .659 &\textcolor{red!45}{$.675$}\\
     HuBERT-large& .514 & .209 & .419 & .604 & \textcolor{red!70}{$.364$} & \textcolor{red!45}{$.567$} & .535 & .428 & .471 & .585 & .581 & .588\\
     wav2vec-large& .464 & .149 & .333 & .475 & .200 & .306 & .470 & .358 & .397 & .583 & .557 &.578\\
     wav2vec2-base& .481 & .143 & .312 & .527 & .253 & .449 & .481 & .333 & .312 & .585 & .581&  .588\\
     wav2vec2-large & .481 & .143 & .312 & .510 & .254 & .415 & .482 & .335 & .315 & .580 & .583 & .584\\
     wav2vec2-large-age-gender& .488 & .235 & .427 & .580 & .315 & .531 & .534 & .412 & .459 & .574 & .591 & .576\\
     wav2vec2-large-xlsr-53& .482 & .144 & .314 & .573 & .250 & .471 & .489 & .344 & .339 & .583 & .557 & .576\\
     WavLM-base& .488 & .208 & .416 & .588 & .332 & .546 & .528 & .443 & .496 & .646 & .613 &.635\\
     WavLM-large& \textcolor{red!45}{$.515$} & \textcolor{red!100}{$.262$} & \textcolor{red!100}{$.470$} & \textcolor{red!100}{$.615$} & \textcolor{red!45}{$.363$} & \textcolor{red!100}{$.576$} & \textcolor{red!100}{$.575$} & \textcolor{red!70}{$.480$} & \textcolor{red!70}{$.539$} & \textcolor{red!70}{$.710$} & \textcolor{red!100}{$.711$} & \textcolor{red!100}{$.712$}\\
     Whisper-base& .514 & \textcolor{red!70}{$.249$} & \textcolor{red!45}{$.462$} & \textcolor{red!70}{$.608$} & .348 & \textcolor{red!45}{$.567$} & \textcolor{red!45}{$.553$} & .454 & .510 & .662 & .662 & .665\\
     Whisper-large& \textcolor{red!70}{$.519$} & \textcolor{red!45}{$.238$} & .443 & \textcolor{red!45}{$.606$} & \textcolor{red!100}{$.375$} & \textcolor{red!70}{$.574$} & \textcolor{red!70}{$.561$} & \textcolor{red!45}{$.458$} & \textcolor{red!45}{$.513$} & .672 & \textcolor{red!70}{$.680$} & \textcolor{red!45}{$.675$}\\
    \bottomrule
    \end{tabular}
    }
\end{table*}

\begin{table*}[htbp] % 使用table*环境创建一个跨越两栏的表格
    \setlength\tabcolsep{14pt}
    \centering
    % \caption{The Short-term task result for the ParaLBench}
    \caption{\reviewtwo{Results of the selected \textit{short-term} paralinguistic tasks (\ie emotion, Arousal, Valence, and Dominance) in ParaLBench over two handcrafted feature sets and 14 acoustic foundation models.}}
    \label{tab: short-term-1}
    \resizebox{\textwidth}{!}{%
    \begin{tabular}{c|cccccc}
    \toprule
    \multirow{2}*{Foundation models}&  \multicolumn{3}{c}{Emotion (IEMOCAP, 4-cls)} &  \multicolumn{1}{c}{Arousal (IEMOCAP)} &  \multicolumn{1}{c}{Valence (IEMOCAP)} &  \multicolumn{1}{c}{Dominance (IEMOCAP)}\\
     &  WA&  UA&  WF1&   MAE&  MAE&  MAE \\
    \midrule
    \midrule
    eGeMAPS & .365& .288& .269& .135& .198& .163\\
     ComParE-2016& .330& .294& .294& .132& .194& .159\\
     \midrule
     CLAP-HTSAT & .459& .483& .451& .109& .186& .145\\
     data2vec-audio-base & .589& .496& .587& .113& .167& .142 \\
     emotion2vec & .629& .629& .625& \textcolor{red!45}{$.099$}& \textcolor{red!45}{$.147$}& .131\\
     HuBERT-base & .598& .602 & .598& \textcolor{red!100}{$.093$}& .155& \textcolor{red!45}{$.130$}\\
     HuBERT-large & \textcolor{red!45}{$.649$}& \textcolor{red!45}{$.643$}& \textcolor{red!45}{$.649$}& .100& .152& .134\\
     wav2vec-large & .429& .425& .427& .119& .191& .144\\
     wav2vec2-base & .451& .475& .452& .124& .184& .147\\
     wav2vec2-large& .449& .451& .442& .120& .183& .145\\
     wav2vec2-large-age-gender&.613& .627& .613& \textcolor{red!100}{$.093$}& .157& .131\\
     wav2vec2-large-xlsr-53 & .525& .563& .503& .103& .184& .133\\
     WavLM-base & .620& .624& .618& \textcolor{red!70}{$.096$}& .154& .132\\
     WavLM-large & \textcolor{red!100}{$.690$}& \textcolor{red!70}{$.685$}& \textcolor{red!100}{$.687$}& .103& \textcolor{red!70}{$.144$}& \textcolor{red!70}{$.128$}\\
     Whisper-base & .634& .634& .632& \textcolor{red!45}{$.099$}& \textcolor{red!70}{$.144$}& \textcolor{red!45}{$.130$}\\
     Whisper-large & \textcolor{red!100}{$.682$}& \textcolor{red!100}{$.690$}& \textcolor{red!70}{$.682$}& \textcolor{red!70}{$.096$}& \textcolor{red!100}{$.141$}& \textcolor{red!100}{$.127$}\\
     
    \bottomrule
    \end{tabular}
    }
\end{table*}

\begin{table*}[htbp] % 使用table*环境创建一个跨越两栏的表格
    \setlength\tabcolsep{8.1pt}
    \centering
    % \caption{The Medium-term task result for ParaLBench}
    \caption{Results of the selected \textit{medium-term} paralinguistic tasks (\ie sarcasm, influenza, and stutter) in ParaLBench over two handcrafted feature sets and 14 acoustic foundation models.}
    \label{tab: medium-term}
    \resizebox{\textwidth}{!}{%
    \begin{tabular}{c|cccccccccccc}
    \toprule
    \multirow{2}*{Foundation models} &  \multicolumn{3}{c}{Sarcasm (MUStARD, 2-class)} &  \multicolumn{3}{c}{Influenza (FluSense, 9-class)} &  \multicolumn{3}{c}{Stutter (SEP-28k, 2-class)}& \multicolumn{3}{c}{\reviewtwo{Depression (DAIC-WOZ, 2-class)}}\\
     &  WA&  UA&  WF1&   WA&  UA&  WF1&  WA&  UA&  WF1&  WA&  UA&  WF1 \\
    \midrule
    \midrule
     eGeMAPS & .435 & .506 & .283 & \editor{.494} & \editor{.214} & \editor{.434} & .551 & .506 & .257& .702& .500& .579\\
     ComParE-2016& .581 & .612 & .568 & \editor{.328} & \editor{.112} & .\editor{172} & .572 & .500 & .416& .702& .500& .579\\
     \midrule
     CLAP-HTSAT & .596 & \textcolor{red!70}{$.610$} & \textcolor{red!45}{$.595$} & .522 & .382 & .513 & .576 & .588 & .570& .723& .536& .626\\
     data2vec-audio-base & \textcolor{black}{.573} & \textcolor{black}{.502} & \textcolor{black}{.427} & .618 & .369 & .589 & .738 & .718 & .731& .702& .500& .579\\
     emotion2vec & .593 & .566 & .579 & .729 & \textcolor{red!45}{$.546$} & .725 & .765 & \textcolor{red!40}{$.770$} & \textcolor{red!45}{$.766$}& \textcolor{red!70}{$.745$}& .592& \textcolor{red!70}{$.691$}\\
     HuBERT-base & \textcolor{red!45}{$.626$} & .599 & \textcolor{red!70}{$.613$} & .621 & \textcolor{red!100}{$.577$} & .600 & .725 & .721 & .725& .723& .577& \textcolor{red!45}{$.674$}\\
     HuBERT-large & \textcolor{black}{.573} & \textcolor{black}{.500} & \textcolor{black}{.417} & .718 & .497 & .704 & \textcolor{red!70}{$.790$} & \textcolor{red!70}{$.785$} & \textcolor{red!70}{$.790$}& .702& .500& .579\\
     wav2vec-large & .559 & .564 & .561 & .515 & .301 & .499 & .622 & .625 & .624& .723& .597& \textcolor{red!70}{$.691$}\\
     wav2vec2-base & \textcolor{black}{.522} & \textcolor{black}{.560} & \textcolor{black}{.495} & .490 & .205 & .431 & .695 & .678 & .690& .702& .500& .579\\
     wav2vec2-large& .427 & .500 & .256 & .480 & .209 & .423 & .695 & .655 & .667& .702& .500& .579\\
     wav2vec2-large-age-gender& .542 & .584 & .509 & .696 & .482 & .665 & .758 & .746 &.756& .723& \textcolor{red!45}{$.597$}& \textcolor{red!70}{$.691$} \\
     wav2vec2-large-xlsr-53 & \textcolor{black}{.573} & \textcolor{black}{.500} & \textcolor{black}{.417} & .523 & .318 & .496 & .629 & .605 & .618& .702& .500& .579\\
     WavLM-base & .565 & .573 & .566 & .664 & .414 & .637 & .748 & .749 & .749& \textcolor{red!100}{$.766$}& \textcolor{red!100}{$.710$}& \textcolor{red!100}{$.763$}\\
     WavLM-large & \textcolor{red!70}{$.629$} & \textcolor{red!45}{$.605$} & \textcolor{red!100}{$.619$} & \textcolor{red!70}{$.753$} & .539 & \textcolor{red!70}{$.740$} & \textcolor{red!100}{$.803$} & \textcolor{red!100}{$.799$} & \textcolor{red!100}{$.803$}& \textcolor{red!70}{$.745$}& .571& .669\\
     Whisper-base & .559 & .583 & .551 & \textcolor{red!45}{$.741$} & \textcolor{red!70}{$.562$} & \textcolor{red!100}{$.733$} & .733 & .725 &.732& .617& \textcolor{red!70}{$.645$}& .633\\
     Whisper-large & \textcolor{red!100}{$.698$} & \textcolor{red!100}{$.618$} & \textcolor{red!45}{$.595$} & \textcolor{red!100}{$.762$} & \textcolor{red!100}{$.577$} & \textcolor{red!100}{$.756$} &  \textcolor{red!45}{$.767$} & .753 & .764& .723& .556& .654\\
    \bottomrule
    \end{tabular}
    }
\end{table*}

\begin{table*}[htbp] % 使用table*环境创建一个跨越两栏的表格
    \vspace{-0.2cm}
    \setlength\tabcolsep{10.2pt}
    \centering
    % \caption{The Long-term task result for ParaLBench}
    \caption{Results of the selected \textit{long-term} paralinguistic tasks (\ie gender, age, accent, and dialect) in ParaLBench over two handcrafted feature sets and 14 acoustic foundation models.}
    \label{tab: long-term}
    \resizebox{\textwidth}{!}{%
    \begin{tabular}{c|ccccccccccccc}
    \toprule
    \multirow{2}*{Foundation models}&  \multicolumn{3}{c}{Gender (MSP-Podcast, 2-class)}  &  \multicolumn{1}{c}{Age (VCTK)} &  \multicolumn{3}{c}{Accent (VCTK, \textcolor{black}{11-class})}  & \multicolumn{3}{c}{Dialect (TIMIT, 8-class)}\\
     &  WA&  UA&  WF1& \textcolor{black}{MAE} & WA&  UA&  WF1&  WA&  UA&  WF1 \\
    \midrule
    \midrule
     eGeMAPS &  .786&   .785&  .786&  6.481&  \textcolor{black}{.362}& \textcolor{black}{.129} & \textcolor{black}{.236} & .178 & .128 &.106\\
     ComParE-2016& .583 & .579 & .568  & 2.425 &  .323& .100 & .158 & \textcolor{black}{.152}  & .123 &.114\\
     \midrule
     CLAP-HTSAT& .876 & .875 & .756  & \textcolor{red!70}{$1.862$} & \textcolor{red!70}{$.984$} & \textcolor{red!100}{$.986$} & \textcolor{red!70}{$.984$} & .199 & .158 &.162\\
     data2vec-audio-base& .815 & .815 & .815 & 1.942 & .576 & .343 & .547 & .186 & .148 &.133\\
     emotion2vec& .931 & .931 & .931 & 1.888 & .946 & .943 & .946 & .206 & .185 &.202\\
     HuBERT-base& \textcolor{red!45}{$.970$} & \textcolor{red!70}{$.970$} & \textcolor{red!45}{$.970$}  & \textcolor{red!45}{$1.867$} & \textcolor{red!100}{$.990$} & \textcolor{red!45}{$.979$} & \textcolor{red!100}{$.990$} & \textcolor{red!45}{$.264$} & \textcolor{red!70}{$.239$} & \textcolor{red!45}{$.252$}\\
     HuBERT-large& .931 & .931 & .931 & 1.951 & .794 & .692 & .783 & .210 & .167 &.154\\
     wav2vec-large& .550 & .545 & .523 & 2.157 & .380 & .164 & .279 & .164& .148 &  .157\\
     wav2vec2-base& .562 & .561 & .561 & 1.946 & .392 & .143 & .289  & .175 & .144 &.128\\
     wav2vec2-large& .644 & .644 & .644  & 1.942 &  .438&  .171&  .328 &  .177& .141 & .087\\
     wav2vec2-large-age-gender& \textcolor{red!100}{$.985$} & \textcolor{red!100}{$.984$}  & \textcolor{red!100}{$.985$} & 2.070 & \textcolor{red!45}{$.979$} & \textcolor{red!70}{$.982$} & \textcolor{red!45}{$.979$} & .210& .171& .181\\
     wav2vec2-large-xlsr-53& .865 & .867 & .864 & 2.021 & .484 & .279 & .436 & .188& .161& .141\\
     WavLM-base& .966 & \textcolor{red!45}{$.966$} & .966 & \textcolor{red!100}{$1.826$} & .937 & .946 & .937 & \textcolor{red!70}{$.271$} &  \textcolor{red!45}{$.237$} & \textcolor{red!70}{$.257$}\\
     WavLM-large& \textcolor{red!70}{$.984$} & \textcolor{red!100}{$.984$}  & \textcolor{red!70}{$.984$} & 2.000 & .969 & .970 & .970 & \textcolor{red!100}{$.295$}& \textcolor{red!100}{$.268$}& \textcolor{red!100}{$.272$}\\
     Whisper-base& .931 & .932 & .931 & 1.984 & .688 & .619 & .673 & .218& .174& .202\\
     Whisper-large & .934 & .935 & .934 & 2.212 & .835 & .795 & .834 & .248 & .192 & .194\\
    \bottomrule
    \end{tabular}
    }
\end{table*}

\subsection{Full Benchmark Results}
\subsubsection{Handcrafted Features vs SSL Acoustic Foundation Models}
Since \textcolor{black}{eGeMAPS} and ComParE-2016 are widely used as baselines for evaluating various tasks in computational paralinguistics, we employ them in ParaLBench for baseline comparisons. According to the results in Tables~\ref{tab: short-term},~\ref{tab: medium-term}, and ~\ref{tab: long-term}, we find that the performance of acoustic foundation models surpasses that of handcrafted features in most tasks. Nonetheless, exceptions are noted. For example, in emotion and sentiment tasks, eGeMAPS performs comparably to, and sometimes even exceeds, some acoustic foundation models. This likely stems from eGeMAPS features being specifically designed for speech emotion recognition. Similarly, ComParE-2016 also achieves performance comparable to acoustic foundation models in specific tasks like gender prediction and sarcasm detection. These findings suggest that specially designed handcrafted features can still compete with acoustic foundation models in certain scenarios.

% We have utilized eGeMAPS and ComParE-2016, prevalent in ComParal, as baseline references for various tasks within ParaLBench. Observations from Tables~\ref{tab: short-term},~\ref{tab: medium-term}, and~\ref{tab: long-term} emphasize that acoustic foundation models generally outperform handcrafted features in most tasks. Nonetheless, exceptions are noted. In emotion and sentiment analysis tasks, eGeMAPS demonstrates competitive, and at times superior, performance compared to some acoustic foundation models, likely due to its specialized focus on speech emotion recognition. Similarly, ComParE-2016 exhibits comparable efficacy to the foundation models in areas like gender prediction and sarcasm detection, suggesting that specialized handcrafted features can rival acoustic models under specific conditions.

Overall, with the increase in training data for speech foundation models and the optimization of their structures, these models typically outperform handcrafted features. However, this superiority does not imply that handcrafted features lack advantages. For instance, handcrafted features demand fewer computational resources compared to acoustic foundation models, a crucial consideration in resource-constrained environments.

\subsubsection{Impact of Training Data Volume on Foundation Models}
% 训练数据的多少，是否经过专门数据的微调
% Deep learning is a data-driven discipline, and generally, model performance improves with increased data volume. This phenomenon has been validated in the performance of several acoustic foundation models. For instance, wav2vec2-large, WavLM-large, and HuBERT-large have similar parameter sizes. As shown in Table~\ref{tab:dataset}, these three foundation models utilize different amounts of training data. Ignoring factors such as model architecture and training methods, WavLM-large, which uses the most training data, achieves the best performance in the majority of tasks. In contrast, wav2vec-large, trained with only 960 hours of Librispeech data, performs worse than handcrafted features in some tasks. This indicates that increasing the amount of training data can significantly enhance the generalization performance of speech foundation models in computational paralinguistic tasks. However, we found that models like wav2vec2-large-xlsr-53, which incorporate additional multilingual training data, might experience negative impacts on tasks that are purely in English.

In deep learning, the axiom that ``there is no data like more data'' often holds true, a notion supported by the results achieved by various acoustic foundation models. Among models with similar capacities, such as wav2vec2-large, WavLM-large, and HuBERT-large, differences in their training data volumes are illuminating. As shown in Table~\ref{tab:dataset}, these models are trained with varying data sizes. Disregarding other variables such as architectural specificities and training methodologies, it is observed that WavLM-large, benefiting from the most extensive training dataset, consistently outshines its peers in most tasks. Conversely, wav2vec2-large suffers in some areas when trained with only 960 hours of Librispeech data—less voluminous than its counterparts—underscoring the significance of data volume in bolstering the generalization capacity of speech foundation models for computational paralinguistic applications. Interestingly, we note instances where additional diversity in training data, such as including multiple languages as in the wav2vec2-large-xlsr-53 model, might detrimentally affect performance on tasks strictly in the English language. This highlights the nuanced interplay between data variety and task specificity.

% Additionally, we observe that further pre-training on task-specific datasets can improve acoustic foundation model performance for those tasks. For example, wav2vec2-large-age-gender, compared to wav2vec2-large, undergoes additional pre-training on datasets related to age and gender, resulting in superior performance in gender and age prediction tasks. Another example is emotion2vec, a foundation model specifically developed for speech emotion recognition. Its training data is primarily emotion-related, and despite having a parameter size comparable to other base models, emotion2vec outperforms most large models in emotion and sentiment tasks.

Furthermore, we observe that further pre-training on task-specific datasets can improve acoustic foundation model performance for those tasks. For example, wav2vec2-large-age-gender, compared to wav2vec2-large, undergoes additional pre-training on datasets related to age and gender, resulting in superior performance in gender and age prediction tasks. Another example is emotion2vec, a foundation model specifically developed for speech emotion recognition. Its training data is primarily emotion-related, and despite having a parameter size comparable to other base models, emotion2vec outperforms most large models in emotion and sentiment tasks.

\subsubsection{Impact of Model Structure on Performance}
% CNN， Transformer encoder， Transformer encoder-decoder， base or large
The impact of model structure on performance is a critical consideration in the development of more general-purpose acoustic foundation models. Researchers have opted for various model architectures, showcasing exceptional performance in several general tasks, including automatic speech recognition~\cite{radford2022whisper, hsu2021hubert}, and speech translation~\cite{radford2022whisper}. However, evidence supporting their effectiveness in a broader range of ComParal tasks is lacking.

% As shown in Table~\ref{foundation model}, we present the architectures used by different acoustic foundation models. Except for wav2vec-large, which uses CNN as its base architecture, the other models all employ Transformer architectures. Comparing wav2vec-large and wav2vec2-large, both of which use similar training methods and the same training data, we observe that wav2vec2-large, based on the Transformer architecture, performs better in the majority of tasks. This indicates that the Transformer architecture has a stronger ability to capture paralinguistic information, likely due to its superior capacity for modeling long-range dependencies in speech. Meanwhile, the performance of the Large model is generally better than that of the base model, which suggests that increasing the number of layers in the Transformer or increasing the number of learnable parameters can lead to improved performance.

Table~\ref{foundation model} provides detailed insights into the architectural foundations of various acoustic foundation models. With the exception of wav2vec-large, which uses a CNN-based architecture, the majority of models predominantly utilize Transformer-based structures. A more in-depth comparative analysis between wav2vec-large and wav2vec2-large models, sharing the same training methodology and data, reveals that the Transformer-incorporating wav2vec2-large model outperforms its CNN counterpart across most paralinguistic challenges. This observation suggests that the Transformer architecture excels in capturing paralinguistic nuances, possibly due to its proficiency in processing long-range sequential dependencies within speech data. Furthermore, models designated as \editor{`large'} consistently outperform their \editor{`base'} counterparts, indicating that architectural enhancements, whether in dimensionality or layer count, can significantly enhance performance.

\subsubsection{Impact of Training Methods on Model Performance}
Different training methods can significantly influence a model's ability to capture various paralinguistic information. For instance, wav2vec 2.0 employs contrastive learning tasks to learn latent representations using quantized speech representations. Conversely, HuBERT and WavLM utilize clustering methods to generate pseudo-labels as training targets. Data2vec and emotion2vec follow a Teacher-Student approach, where the Teacher provides training targets for the Student. CLAP aligns text and speech representations through contrastive learning, while Whisper employs an autoregressive learning method to learn latent representations and undergoes supervised training on multiple tasks.

% Comparing the results in the table, we find that HuBERT and WavLM, which use clustering methods, exhibit the best generalization performance across multiple tasks. WavLM, in particular, shows superior generalization due to the incorporation of noise during training. Whisper, trained using an autoregressive method and supervised on multiple tasks, also demonstrates strong generalization capabilities, achieving the best performance in tasks such as Sarcasm and Influenza detection. CLAP, which uses contrastive learning, shows stable results across various tasks. In contrast, data2vec, emotion2vec, and wav2vec 2.0 exhibit relatively mediocre performance in multiple computational paralinguistic tasks.

Comparing the results in all the tables mentioned above, we observe that HuBERT and WavLM, which employ clustering-based methods (e.g., spectral clustering, hierarchical clustering), demonstrate the best generalization performance across multiple paralinguistic tasks. WavLM's exceptional performance in generalization can be partly attributed to its exposure to noise during the training phase. Similarly, Whisper, trained using an autoregressive method and supervised on multiple tasks, also exhibits strong generalization capabilities, achieving the best performance in tasks such as Sarcasm and Influenza detection. CLAP, which uses contrastive learning, shows stable results across various tasks. In contrast, data2vec, emotion2vec, and wav2vec 2.0 performed less competitively across a wider range of tasks.

\subsection{Acoustic Foundation Model on Different \reviewtwo{Terms}}
% 对于不同的SSL模型在不同Term上的表现不同
As shown in Table~\ref{tab: short-term},~\ref{tab: medium-term}, and~\ref{tab: long-term}, we present the results of various tasks along the time axis, from short-term states to long-term traits. These findings delineate the inclinations of the acoustic foundation models across tasks situated at distinct junctures on the timeline, offering guidance for subsequent researchers in the selection of speech foundation models.

% \subsubsection{Performance Analysis of Acoustic Models on Short-term Tasks}
\subsubsection{Short-Term Task Performance Analysis}

% For some states that can be entered in a Short time, we define them as short-term. In Table ~\ref{tab: short-term}, we present two short-term tasks: Emotion and Sentiment. Acoustic foundation models learn latent paralinguistic information from a large amount of data, which may include emotion-related information, but the performance of different foundation models varies in this regard. In Table~\ref{tab: short-term}, we can see that the two models with the best generalization performance in Short-term tasks are emotion2vec and WavLM-large. The former is specifically optimized for emotion, including datasets and model architectures, which helps the model learn latent paralinguistic information. The latter, compared to HuBERT, incorporates more training data and introduces noise, allowing the model to learn more paralinguistic information, including emotion, and thus demonstrate good performance in Short-term tasks.  Interestingly, we found that the wav2vec2-large-age-gender model, which underwent additional training on gender data, exhibited significantly improved performance in Emotion (MSP-Podcast) and Sentiment (MELD) compared to other versions of wav2vec 2.0. This suggests a certain correlation between long-term tasks such as gender prediction and short-term tasks.

In the analysis of short-term task performance, we focus on states that manifest over a brief duration. Tables~\ref{tab: short-term} reflect the model performance in two such tasks: Emotion and Sentiment. While acoustic foundation models handle extensive datasets potentially containing emotion-specific information, their efficacy varies considerably. Notably, emotion2vec and WavLM-large demonstrate exceptional performance in short-term tasks. Emotion2vec benefits from tailored datasets and architecture, enhancing its proficiency in emotion-centric paralinguistic learning, whereas WavLM-large, leveraging its extensive dataset and noise-augmented training regimen, excels in capturing transient states. Interestingly, wav2vec2-large-age-gender, through its gender-focused further training, exhibits notable enhancements in the Emotion (MSP-Podcast) and Sentiment (MELD) tasks, indicating a correlation between learning long-term attributes like gender and the recognition of short-term emotional states.

% \subsubsection{Performance Analysis of Acoustic Models on Medium-term Tasks}
\subsubsection{Medium-Term Task Performance Analysis}
In the analysis of medium-term task performance, we consider states that typically manifest between long-term and short-term durations, encompassing partially self-induced or temporary states, as well as potential anomalous speech. Within Table~\ref{tab: long-term}, we investigate three medium-term tasks: Sarcasm, Influenza, and Stutter. The Whisper-large model excels in the Sarcasm and Influenza tasks, likely due to its Encoder-Decoder Transformer design and skills honed through autoregressive training, features that align well with the requirements of medium-term tasks. Similarly, the WavLM-large model demonstrates its effectiveness in Stutter detection.

% \subsubsection{Performance Analysis of Acoustic Models on Long-term Tasks}
\subsubsection{Long-Term Task Performance Analysis}
% In our analysis of long-term task performance, we designate traits that humans develop over extended periods as long-term traits. Within Table~\ref{tab: long-term}, we primarily present four tasks: Gender, Age, Accent, and Dialect. Surprisingly, we observed that in long-term tasks, some base foundation models can even match or surpass large foundation models. For instance, in the Age task, the best-performing model is WavLM-base, not WavLM-large. Similarly, in the Accent and Dialect tasks, HuuBERT-base outperforms HuBERT-large. Additionally, wav2vec2-base performs comparably to wav2vec2-large in the Age and Dialect tasks. This suggests that relatively smaller base models may be more suitable for long-term tasks. Moreover, we noted that CLAP performs well in the Gender, Age, and Accent tasks. This could be attributed to the contrastive learning approach employed by CLAP, which compares positive and negative samples, enabling the model to learn more discriminative feature representations in long-term tasks, thus enhancing its performance in classification and clustering tasks.

In our analysis of long-term task performance, we designate traits that humans develop over extended periods as long-term traits. Table~\ref{tab: long-term} primarily focuses on four tasks: Gender, Age, Accent, and Dialect. Interestingly, we observed that in these long-term tasks, some base models can even match or surpass large foundation models. Contrary to expectations, the best-performing model in the Age task is WavLM-base, not WavLM-large. Similarly, HuuBERT-base outperforms HuBERT-large in the Accent and Dialect tasks. Additionally, wav2vec2-base performs comparably to wav2vec2-large in the Age and Dialect tasks. This suggests that relatively smaller base models may be more suitable for long-term tasks.

% 等结果出来分析下结果
% 途中画出本数据集到本数据集的数值，画一条线
\subsection{Cross-corpus Analysis}
\begin{figure*}[!t]
  \vspace{-0.5cm}
  \centering
  \includegraphics[width=0.9\textwidth, trim=0 10 0 0, clip]{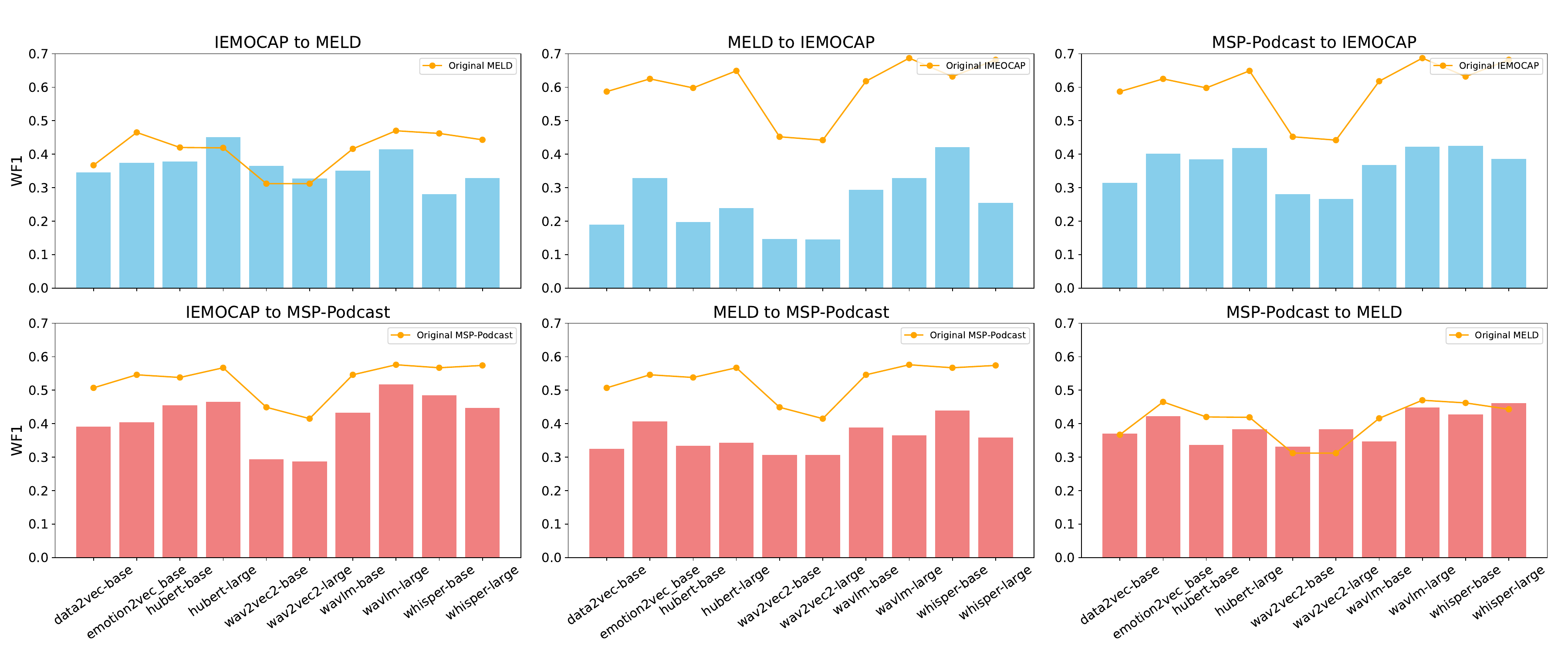}
  \vspace{-0.2cm}
  \caption{\reviewtwo{Cross-corpus results of the acoustic foundation model on three emotion datasets: IEMOCAP, MELD, and MSP-Podcast. The subtitles, such as 'IEMOCAP to MELD,' indicate that the model is trained on the IEMOCAP dataset and tested for cross-corpus performance on the MELD dataset. The orange line represents the results achieved by the models on the original dataset.}}
  \label{fig:cross_dataset}
\end{figure*}

\reviewtwo{This section aims to explore the generalization performance of acoustic foundation models in cross-corpus. Given the similarity of emotion labels, we selected the IEMOCAP, MELD, and MSP-Podcast datasets for cross-corpus experiments. However, these datasets differ in label categories, such as IEMOCAP (happiness, sadness, anger \& excited, neutral), MSP-Podcast (anger, neutral, happiness, sadness, disgust), and MELD (anger, neutral, joy, sadness, disgust, surprise, fear). To address this issue, we selectively filter out certain labels from the test datasets. For instance, when testing the cross-corpus performance of a model trained on IEMOCAP using the MSP-Podcast dataset, we exclude instances labeled as disgust. Similarly, we remove instances labeled as disgust, surprise, and fear from the MELD dataset.}

\reviewtwo{Figure~\ref{fig:cross_dataset} presents the results of the speech foundation models in cross-corpus experiments. Due to space limitations, we selected ten different speech foundation models to showcase their cross-corpus performance. As shown in Figure~\ref{fig:cross_dataset}, emotion2vec-base, WavLM-base, and WavLM-large exhibit the best performance across multiple cross-corpus experiments, while wav2vec2-base and wav2vec2-large perform poorly in several experiments. By analyzing the performance of these models in Tables~\ref{tab: short-term} and~\ref{tab: short-term-1}, we observe that models with strong performance on within-corpus tend to perform better in cross-corpus tasks, while those with weaker within-corpus performance tend to struggle. Another notable observation is that base models generally underperform compared to large models with more parameters in cross-corpus tasks.}

% 所有数据集全做
\subsection{Efficient fine-tuning using LoRA}
\begin{figure*}[!t]
  \vspace{-0.5cm}
  \centering
  \includegraphics[width=0.9\textwidth, trim= 0 28 0 0, clip]{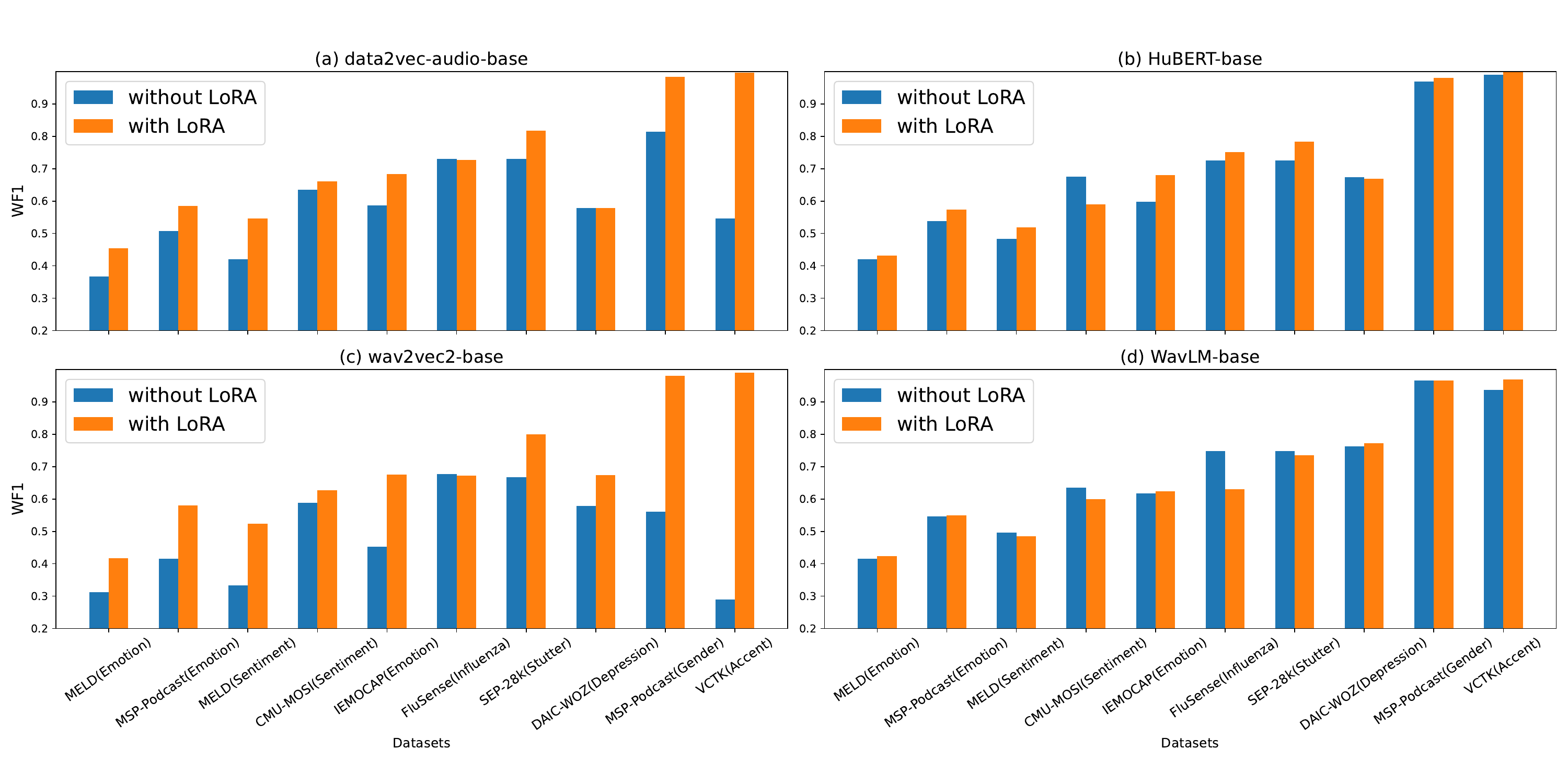}
  \vspace{-0.3cm}
  %\caption{\editor{Efficient fine-tuning of the acoustic foundation models using LoRA.}}
  \caption{\textcolor{black}{Performance comparison of acoustic foundation models with and without LoRA efficient fine-tuning.}}
  \label{fig:lora}
\end{figure*}

% \editor{LoRA~\cite{hu2021lora} (Low-Rank Adaptation) is an efficient fine-tuning technique that offers several significant advantages over full fine-tuning. Firstly, LoRA introduces low-rank matrices for parameter adjustment, drastically reducing the number of parameters that need updating during the fine-tuning process and significantly lowering the computational and storage requirements. This is especially important for fine-tuning large speech foundation models, making the process more lightweight and scalable. Additionally, since LoRA only fine-tunes a subset of parameters, it allows the model to retain the capabilities learned during pre-training while focusing on the new task, thus reducing the risk of overfitting. Overall, LoRA provides an effective alternative to full fine-tuning by efficiently leveraging parameters, cutting computational costs, and maintaining model performance.}

% \editor{In this section, we apply LoRA to acoustic foundation models, freezing the original model parameters and only updating the low-rank matrices within LoRA. For compatibility and cost considerations, we tested four foundation models: data2vec-audio-base, HuBERT-base, wav2vec2-base, and WavLM-base. The results are shown in Figure~\ref{fig:lora}.}

\editor{LoRA~\cite{hu2021lora} (Low-Rank Adaptation) is an efficient fine-tuning technique. By introducing low-rank matrices for parameter adjustments, LoRA significantly reduces the number of parameters that need to be updated, thereby lowering the computational resources and storage requirements, especially when fine-tuning acoustic foundation models. Furthermore, LoRA only fine-tunes a subset of parameters, allowing it to maintain the capabilities of the pre-trained model while focusing on learning new tasks. Therefore, LoRA serves as an effective alternative to full fine-tuning.}

\editor{In this section, we apply LoRA to acoustic foundation models, freezing the parameters of the foundation model and updating only the low-rank matrices of LoRA. For compatibility and cost considerations, we tested four base models: data2vec-audio-base, HuBERT-base, wav2vec2-base, and WavLM-base. The results are shown in Figure~\ref{fig:lora}. The experimental results indicate that after introducing LoRA, most speech foundation models show significant performance improvements in downstream tasks; however, WavLM-base experiences performance degradation in most tasks. This degradation may arise from several factors: first, the parameters of WavLM-base have already been thoroughly trained on large-scale data, so LoRA may not effectively capture the existing features and could introduce interference instead; second, the low-rank matrices of LoRA may not be well-suited to the specific architecture of WavLM-base, resulting in the loss of important information.}

% 只做wavlm-base与large版本，所有数据集都做
\subsection{The effect of hidden states of each layer}
\begin{figure*}[htb]
  \vspace{-0.3cm}
  \centering
  \includegraphics[width=0.9\textwidth, trim= 0 10 0 0, clip]{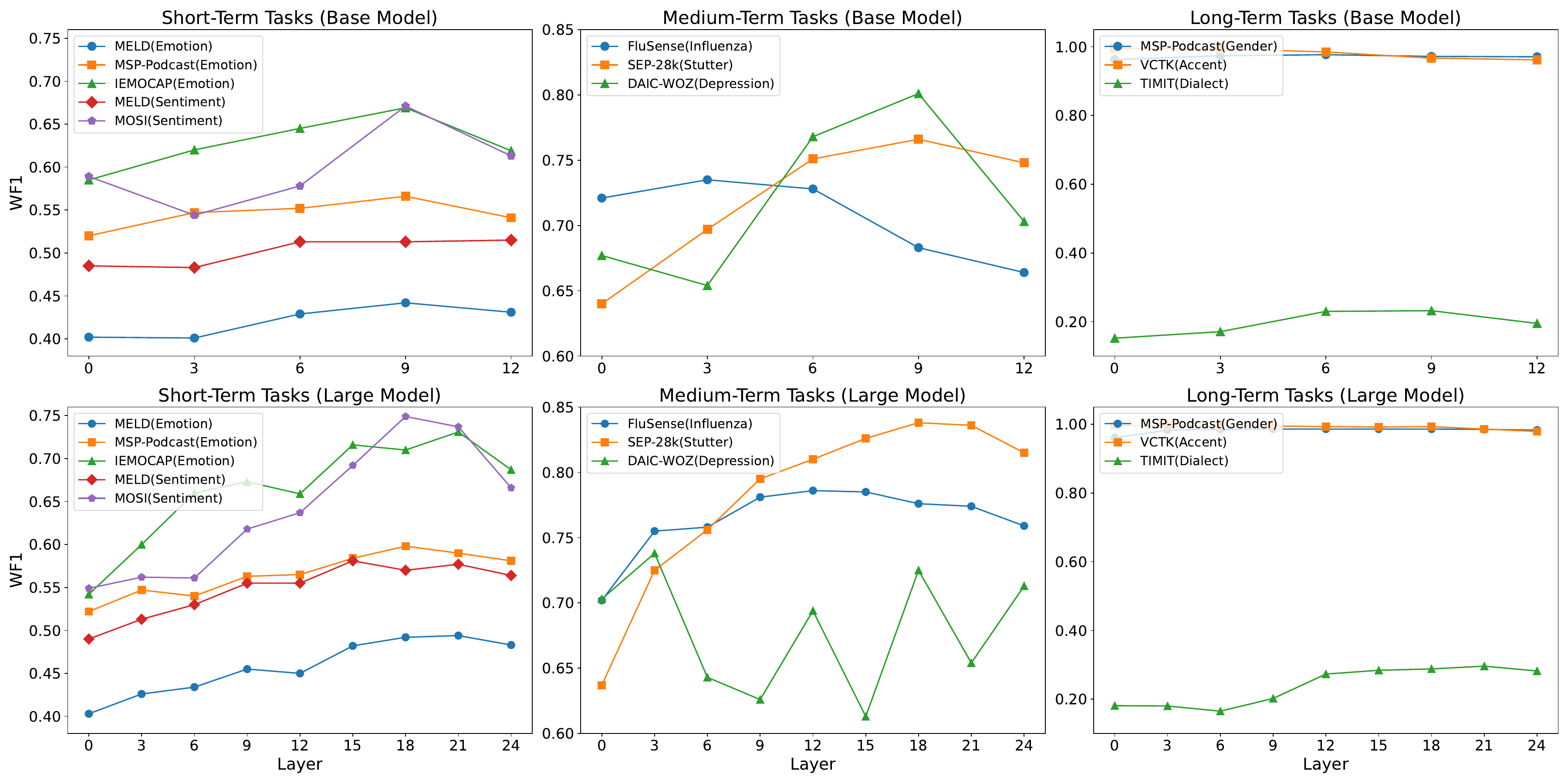}
  \vspace{-0.2cm}
  \caption{\reviewtwo{The performance of the WavLM model layer feature across short, medium, and long-term datasets. The upper and lower figures display the Base and Large versions of the acoustic foundation model. Considering the computational cost, we test the results every 3 layers apart.}}
  \label{fig:layer}
\end{figure*}

\reviewtwo{The architecture of most contemporary acoustic foundation models consists of multiple stacked Transformer layers. For example, the WavLM-base model has 12 layers, while the WavLM-Large model has 24. Research~\cite{DBLP:conf/rep4nlp/TurtonSV21, DBLP:journals/frai/IlinykhD21} has shown that different Transformer layers capture varying levels of semantic information, making it valuable to analyze their performance across paralinguistic tasks. Given that ParaLBench includes numerous such tasks and considering computational constraints, we selected the WavLM model, which has shown strong performance across multiple paralinguistic tasks, for testing. The results in \textcolor{black}{Figure}~\ref{fig:layer} illustrate the performance of different Transformer layers across various tasks, offering insights for optimizing speech foundation models in the future.}

\reviewtwo{The results indicate that, whether in short, medium, or long-term paralinguistic tasks, the best performance does not come from the model's last hidden state but rather from the last few layers of the model. This trend is consistent in both the WavLM-base and WavLM-large versions, suggesting that deep hidden states in the model tend to capture more critical semantic features across different tasks and model sizes. This finding not only corroborates the theory that Transformer layers extract distinct features but also emphasizes the importance of leveraging the internal multi-layer features of the model in practical applications. In the future, focusing more on these intermediate layers during fine-tuning and feature extraction may significantly enhance the model's generalization capability in a multi-task environment.}

% \subsection{Future Works}
% Considering that our work aims to establish a large-scale computational paralinguistics benchmark based on acoustic foundation models, we plan to incorporate more computational paralinguistics datasets and acoustic foundation models in future work to enhance our benchmark. Additionally, we will conduct further experiments to verify the robustness of the foundation models in computational paralinguistics. For instance, we will perform experiments on multilingual datasets to validate the models' performance in multilingual environments, analyze the paralinguistic information contained in different layers of the foundation models, and explore how multitask learning can facilitate mutual improvements across different tasks.

\subsection{Last Hidden State vs. Fusion State}

\begin{table}[!t]
  \vspace{-0.2cm}
  \setlength\tabcolsep{5pt}
  \centering
  % \caption{\editor{Last Hidden State vs. Fusion State}}
  \caption{\textcolor{black}{Evaluation of Last Hidden State vs. Fusion State Representations in Downstream Tasks}}
  \label{tab:fusion State}
  \begin{tabular}{cc|ccc|ccc}
    \toprule
    \multirow{2}*{Dataset} & \multirow{2}*{Tasks} & \multicolumn{3}{c}{Last Hidden State} & \multicolumn{3}{c}{Fusion State}\\
     &  &  WA&  UA&  WF1&  WA&  UA&  WF1\\
    \midrule
    \midrule
    MELD & Emotion & .515 & .262 & .470 & .538& .278& .488\\
    MSP-Podcast & Emotion & .615& .363& .576& .640& .409& .608   \\
    IEMOCAP & Emotion & .690& .685& .687& .706& .727& .704 \\
    MELD & Sentiment & .575& .480& .539& .590& .517& .570\\
    CMU-MOSI & Sentiment & .710& .711& .712& .767& .760& .767  \\
    Flusense & influenza & .753& .539& .740& .781 & .555& .768\\
    SEP-28k & Stutter & .803& .799& .803& .843	& .836	& .842 \\
    DAIC-WOZ & Depression & .745& .571& .669& .723& .577& .674\\
    MSP-Podcast & Gender & .984& .984& .984& .988& .987& .988 \\
    VCTK & Accent & .969& .970& .970& .983& .982& .983 \\
    TIMIT & Dialect & .295 & .268 & .272& .248 & .222& .147 \\
    \bottomrule
  \end{tabular}
\end{table}

%Considering that relying solely on the last hidden state may overlook critical task-related information, we adopted a learnable weighted-sum method to fuse~\cite{yang2024large} the hidden states from all layers into a unified representation, as detailed in Table~\ref{tab:fusion State}. The experimental results demonstrate that the Fusion State consistently outperforms the Last Hidden State in the majority of emotion recognition tasks. For instance, in the IEMOCAP dataset, the Fusion State achieved a weighted accuracy (WA) of $0.706$, compared to $0.690$ for the Last Hidden State. However, it is important to note that not all tasks exhibited the advantages of the Fusion State. This indicates that in certain specific tasks or datasets, the Last Hidden State may retain distinct advantages or application potential. Overall, the findings suggest that the Fusion State is more adept at capturing and integrating information, thereby enhancing the model's overall performance in downstream tasks.

\editor{Considering that using only the last hidden state may overlook some task-relevant key information, we employed a learnable weighted-sum method~\cite{yang2024large, yang2021superb} to fuse the hidden states from all layers into a single representation. The experimental results, as shown in Table~\ref{tab:fusion State}, demonstrate that in most tasks, the fusion state generally outperforms the last hidden state. This indicates that fusion state from multiple layers \textcolor{black}{helps} capture and integrate richer task-relevant semantic information, thereby enhancing the model's overall performance in downstream tasks.
However, it is worth noting that the fusion state does not always provide an advantage in every task. For example, in the DAI-WOZ and TIMIT datasets, the performance of the last hidden state surpasses that of the fusion states. This could be due to the potential introduction of noise when fusing multiple layers of hidden states in certain specific tasks or datasets, leading to suboptimal performance in downstream tasks. Therefore, the last hidden state still holds distinct advantages or value in these scenarios.}

% 后面的实验来不及做
% \subsection{Cross-linguistic Performance Analysis}

% \subsection{Cross-corpus Performance Analysis}

% \subsection{The Influence with each task in Multi-task learning}
% Some tasks guide another task's performance well

% \subsection{Necessity of Fine-tuning the Pretrain Model}

% \subsection{Feature from foundation model each layer}

\section{Conclusion and Future Works}
\label{sec:Conclusion}

In this work, we introduce ParaLBench, a comprehensive benchmark for Computational Paralinguistics (ComParal) evaluation of diverse acoustic foundation models. This extensive benchmark spans \textcolor{black}{thirteen} distinct paralinguistic tasks across \textcolor{black}{ten} datasets and considers 14 widely recognized models in the speech processing domain. By establishing uniform testing criteria, ParaLBench aims to provide a fair and standardized means to evaluate the generalization, efficiency, and robustness of these models across various tasks.
Our comprehensive experiments and subsequent analysis indicate the superiority of self-supervised learning-based speech foundation models over traditional handcrafted feature approaches in terms of both performance and generalizability. Additionally, the study demonstrates the influence of model architecture and data diversity on a model's effectiveness in capturing paralinguistic information.
We further analyze the performance trends of speech foundation models relative to short-term, medium-term, and long-term tasks, delineating paralinguistic tasks along a temporal continuum. With the intent to spur further research and enable uniform evaluations, we will make our code publicly available.

% In our ongoing effort to refine and broaden the scope of our large-scale benchmark for ComParal, guided by acoustic foundation models, future work will entail the incorporation of a wider range of ComParal datasets and additional foundation models. We are committed to conducting extensive experimental evaluations to assess and consolidate the robustness of these models within the ComParal domain. Planned endeavors include conducting evaluations on multilingual datasets to evaluate model adaptability and proficiency across diverse linguistic contexts. Additionally, we intend to explore the intricate network of paralinguistic signals at different model layers, thereby unraveling the nuanced representations involved. Furthermore, we will explore the potential synergies of multitask learning approaches, aiming to uncover how simultaneous training on diverse tasks can lead to mutually beneficial enhancements for the models involved.

In the future, we will expand the benchmark to include more ComParal datasets and foundation models, enhancing its generalizability and usefulness. Additionally, we will conduct multilingual evaluations to assess model performance across diverse languages and \reviewtwo{incorporate multimodal} and \reviewthree{large language models} into our research. These endeavors will not only strengthen ParaLBench but also provide valuable insights into the realm of ComParal and the capabilities of acoustic foundation models.

% \section*{Acknowledgment}
% The work leading to this research was supported by the Guangdong Basic and Applied Basic Research Foundation under Grant Number 2024A1515010112. 

\bibliographystyle{IEEEtran}
\bibliography{refs}

\vspace{-.1cm}
\begin{IEEEbiography}[{\includegraphics[width=1in,height=1.25in,clip,keepaspectratio]{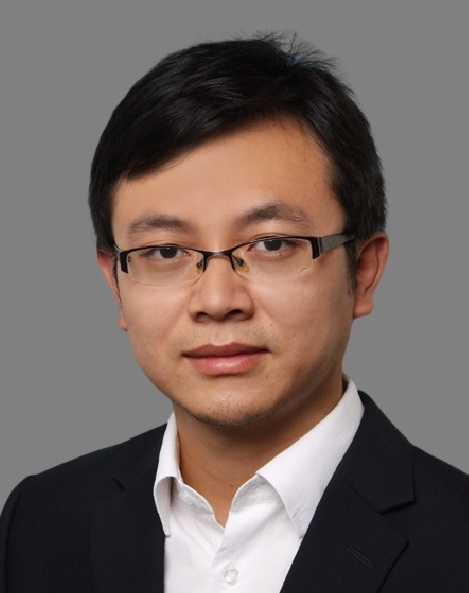}}]
{\bf Zixing Zhang} (M'15-SM'22) received his master degree in physical electronics from the Beijing University of Posts and Telecommunications (BUPT), China, in 2010, and his PhD degree in computer engineering from the Technical University of Munich (TUM), Germany, in 2015. He is now a full professor at the College of Computer Science and Electronic Engineering, Hunan University, China. From 2017 to 2019, he was a research associate with the Department of Computing at the Imperial College London (ICL), UK. Before that, he was a postdoctoral researcher at the University of Passau, Germany. His research focuses on human-centred emotion and health computation. To date, he has authored more than 120 publications in peer-reviewed books, journals, and conference proceedings, leading to more than 5\,500 citations (h-index 44). He serves as an associate editor of the IEEE Transactions on Affective Computing and the Frontiers in Signal Processing, an editorial board member of the Nature Scientific Reports, and a guest editor of the IEEE Transactions on Emerging Topics in Computational Intelligence. 
\end{IEEEbiography}

\vspace{-.7cm}
\begin{IEEEbiography}[{\includegraphics[width=1in,height=1.25in,clip,keepaspectratio]{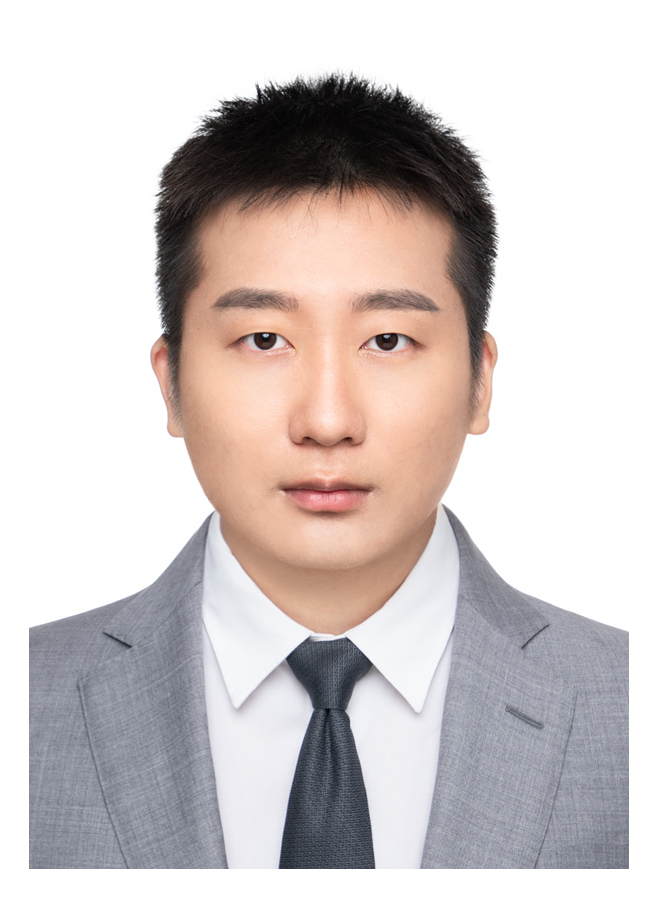}}]{\bf Weixiang Xu} received his master degree from Hunan Normal University, Changsha, China, in 2022. 
He is currently a PhD student at the College of Computer Science and Electronic Engineering, Hunan University, China.  
His research interests include affective computing, multimodal deep learning, and large language models.
\end{IEEEbiography}

\vspace{-.7cm}
\begin{IEEEbiography}[{\includegraphics[width=1in,height=1.25in,clip,keepaspectratio]{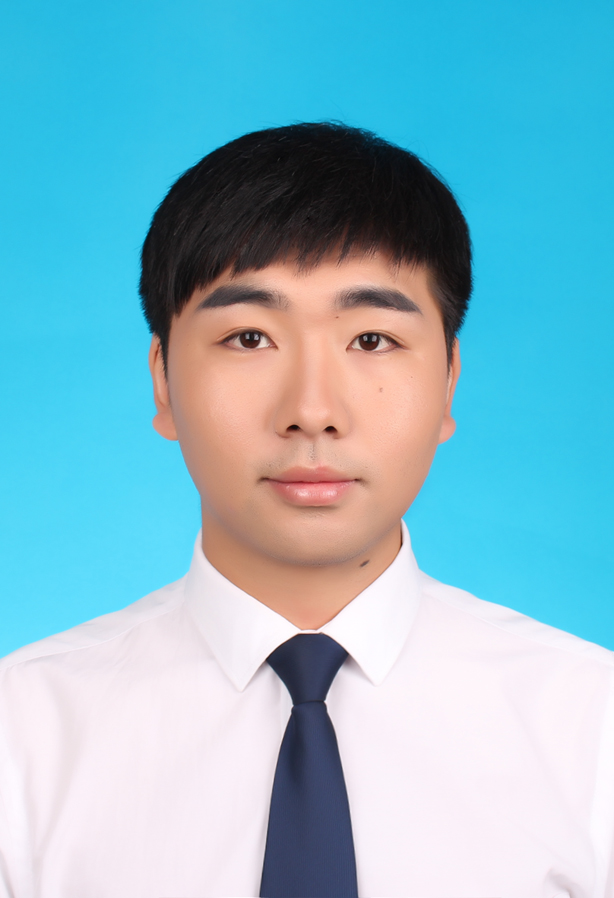}}]{\bf Zhongren Dong} received his master degree from Zhengzhou University, Zhengzhou, China, in 2021.
He is currently a PhD student at the College of Computer Science and Electronic Engineering, Hunan University, China.  
His research interests include self-supervised learning and model compression for mental health and emotion recognition with speech signals.
\end{IEEEbiography}

\vspace{-.7cm}
\begin{IEEEbiography}[{\includegraphics[width=1in,height=1.25in,clip,keepaspectratio]{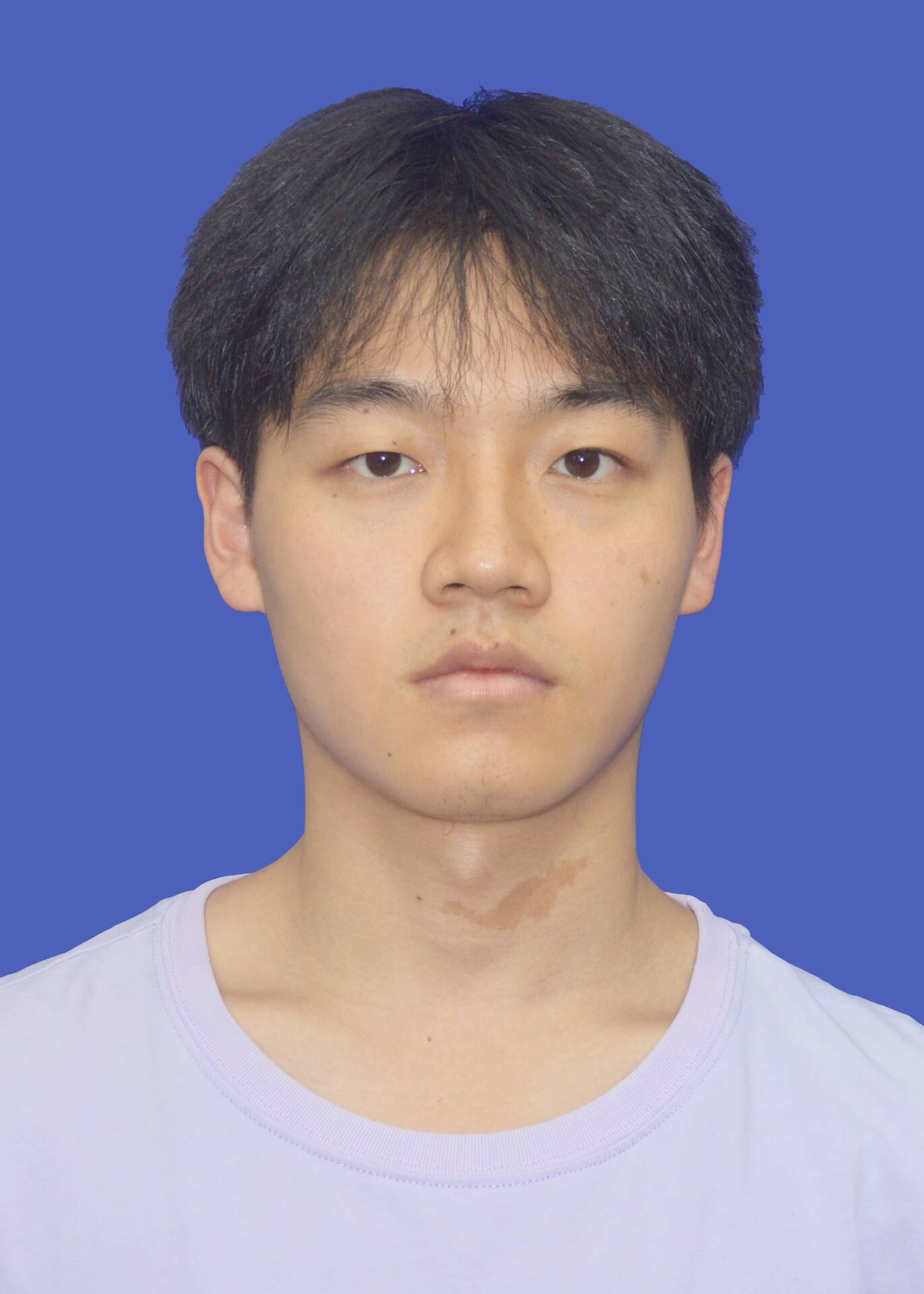}}]{\bf Kanglin Wang} is currently an undergraduate student enrolled in the School of Information Science and Engineering at Hunan University. His research interests focus on computational paralinguistics. %include sublanguage processing, natural language processing, and large language models.
\end{IEEEbiography}

\vspace{-.7cm}
\begin{IEEEbiography}[{\includegraphics[width=1in,height=1.25in,clip,keepaspectratio]{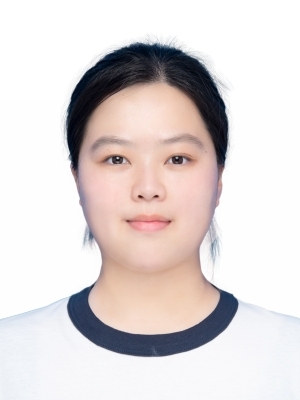}}]{\bf Yimeng Wu} received  the bachelo degree in software engineering from Nanchang University(NCU), Nanchang, China, in 2022. She is a postgraduate student at the College of Computer Science and Electronic Engineering, Hunan University. Her research interests center on the identification of paralinguistic information in speech.
\end{IEEEbiography}

\vspace{-.7cm}
\begin{IEEEbiography}[{\includegraphics[width=1in,height=1.25in,clip,keepaspectratio]{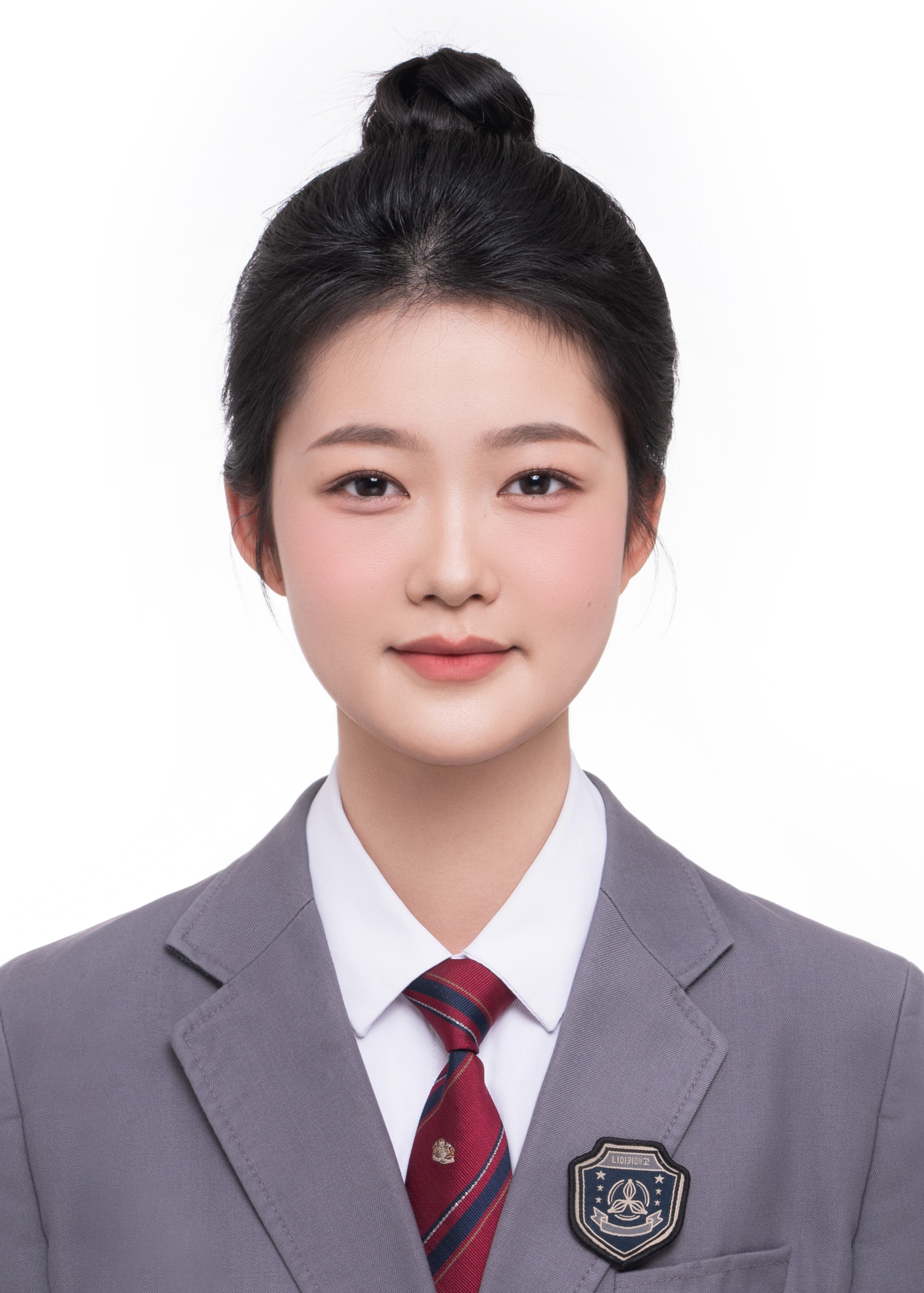}}]{\bf Jing Peng} is currently an undergraduate student enrolled in the School of Information Science and Engineering at Hunan University. Her research interests focus on computational paralinguistics. %Her research interests include sublanguage processing, natural language processing, and large language models.
\end{IEEEbiography}

\vspace{-.7cm}
\begin{IEEEbiography}[{\includegraphics[width=1in,height=1.25in,clip,keepaspectratio]{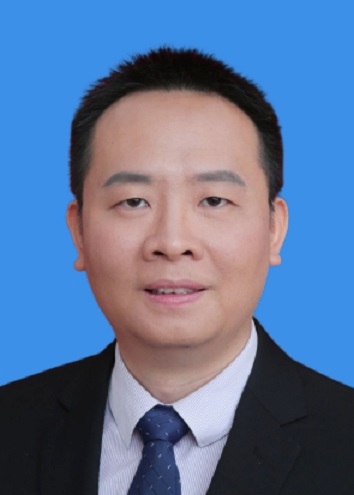}}]{\bf Runming Wang} received the bachelor and master degrees in electronic information engineering from Hunan Normal University (HNNU), in 2003 and 2007, respectively, and the PhD degree from the Huazhong University of Science and Technology (HUST), in 2015. He is currently an Associate Professor with the School of Information Science and Engineering. His current research interests include object detection and recognition, and document image analysis and recognition.
\end{IEEEbiography}

\vspace{-.7cm}
\begin{IEEEbiography}[{\includegraphics[width=1in,height=1.25in,clip,keepaspectratio]{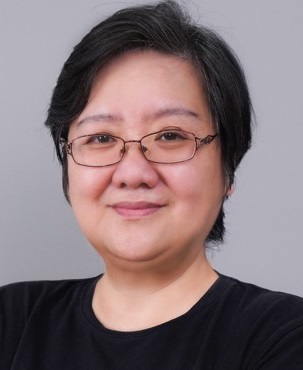}}]
{\bf Dong-Yan Huang} (M'96-SM'05)  received her bachelor and master degrees from  Xi'an  Jiaotong   University,  Xi’an,  China,  in   1985  and   1988, respectively, and the PhD degree in Système Physique et Métrologie-Communication \& Electronique from the  Conservatoire National des Arts et Métiers Paris (CNAM), France, in  1996. 
She is now a Principal Scientist at UBTECH Robotics Corp.  From Dec.~1997 to Nov.~2002, she was a Senior Research Engineer at the Institute of Microelectronics, Singapore.  Before that, she was a postdoctoral researcher at the UFR de Mathématiqueset Informatique, Université Paris Descartes,  France. From Dec. 2002 to 2019, she was a Senior Scientist at the Institute for Infocomm Research, Singapore. Her research focuses on machine learning, pattern recognition,  affective computing, automatic speech recognition, text-to-speech synthesis, voice conversion, computer vision, dialogue system, talking head, human-machine interaction, robotics and embodiment intelligence. She has authored more than 100 publications in peer-reviewed journals, and conference proceedings.
She was solicited and co-chaired for ASMMC from 2015 to 2021. She has been serving as the program committee for several international conferences in the areas of signal processing, speech processing, 
multimedia, human-computer interaction, affective computing and intelligent interaction. She was the chair of the IEEE Singapore Sensor Committee Sub-Committee (2016-2018), the chair of the WIE 
(Women in Engineering) group (2006-2008). She led a team to implement online and offline speech technology on Cruzr, Walker robots and a series of educational products. Her team's works on digital emotion won the awards of A*STAR's 30 Most Impactful Innovations \& Inventions over Three Decades (2021),  and P\&G Connect + Develop Open Innovation Solutions Award on Digital Insights 2020, the first prize in the 2011 INTERSPEECH Speaker State Challenge Sleep Competition and the first prize in the EmotioNet Challenge.
\end{IEEEbiography}

\vfill

\end{document}